 \documentclass[preprint,12pt,3p]{elsarticle}
\graphicspath{ {./figures/} }
\usepackage{subcaption}
\usepackage{amsmath,amsfonts}
\usepackage{algorithm}
\usepackage{algcompatible}
\usepackage[colorlinks = true]{hyperref}
\usepackage{float}
\usepackage[justification=centering]{caption}
\usepackage{verbatim} 
\usepackage{apalike}
\usepackage{array}
\usepackage{amssymb}
\usepackage[
singlelinecheck=false 
]{caption}
\usepackage[usenames,dvipsnames]{color} 
\usepackage[english]{babel}
\addto\extrasenglish{

}  

\DeclareUnicodeCharacter{0307}{\.} 

\usepackage{booktabs}
\usepackage[utf8]{inputenc}
\journal{Engineering Applications of Artificial Intelligence}
\biboptions{authoryear}
\begin{document}
\begin{frontmatter}
\begin{titlepage}
\begin{center}
\vspace*{1cm}

\textbf{ \large A Hybrid Filtering for Micro-video Hashtag Recommendation using Graph-based Deep
Neural Network}
\vspace{1.5cm}

Shubhi Bansal$^{a}$ (phd2001201007@iiti.ac.in), Kushaan Gowda$^b$ (kg3081@columbia.edu), 
Mohammad Zia Ur Rehman$^a$ (phd2101201005@iiti.ac.in),
Chandravardhan Singh Raghaw $^a$(phd2201101016@iiti.ac.in),
Nagendra Kumar$^a$ (nagendra@iiti.ac.in) \\

\hspace{10pt}

\begin{flushleft}
\small  
$^a$ Department of Computer Science and Engineering, Indian Institute of Technology, Indore, India\\
$^b$ Department of Computer Science, Columbia University, New York, USA\\


\vspace{1cm}
\textbf{Corresponding Author:} \\
Shubhi Bansal \\
Department of Computer Science and Engineering,
Indian Institute of Technology, Indore, India\\
Tel: +91-7988361678 \\
Email: phd2001201007@iiti.ac.in

\end{flushleft}        
\end{center}
\end{titlepage}
\title{A Hybrid Filtering for Micro-video Hashtag Recommendation using Graph-based Deep Neural Network}

\author[label1]{Shubhi Bansal\corref{cor1}}
\ead{phd2001201007@iiti.ac.in}
\author[label2]{Kushaan Gowda}
\ead{kg3081@columbia.edu}
\author[label1]{Mohammad Zia Ur Rehman}
\ead{phd2101201005@iiti.ac.in}
\author[label1]{Chandravardhan Singh Raghaw}
\ead{phd2201101016@iiti.ac.in}
\author[label1]{Nagendra Kumar}
\ead{nagendra@iiti.ac.in}
\cortext[cor1]{Corresponding author.}
\address[label1]{Department of Computer Science and Engineering,
Indian Institute of Technology, Indore, India}
\address[label2]{Department of Computer Science, Columbia University, New York, USA}
\begin{abstract}
Due to the growing volume of user-generated content, hashtags are employed as topic indicators to manage content efficiently on social media platforms. However, finding these vital topics is challenging in micro-videos since they contain substantial information in a short duration. Existing methods that recommend hashtags for micro-videos primarily focus on content and personalization while disregarding relatedness among users. Moreover, the cold-start user issue prevails in hashtag recommendation systems. Considering the above, we propose a hybrid filtering-based MIcro-video haSHtag recommendatiON (MISHON) technique to recommend hashtags for micro-videos. Besides content-based filtering, we employ user-based collaborative filtering to enhance recommendations. Since hashtags reflect users’ topical interests, we find similar users based on historical tagging behavior to model user-relatedness. We employ a graph-based deep neural network to model user-to-user, modality-to-modality, and user-to-modality interactions. We then use refined modality-specific and user representations to recommend pertinent hashtags for micro-videos. 
The empirical results on three real-world datasets demonstrate that MISHON attains a comparative enhancement of 3.6\%, 2.8\%, and 6.5\% concerning the F1-score, respectively. Since cold-start users exist whose historical tagging information is unavailable, we also propose a content and social influence-based technique to model the relatedness of cold-start users with influential users. The proposed solution shows a relative improvement of 15.8\% in the F1-score over its content-only counterpart. These results show that the proposed framework mitigates the cold-start user problem.
\end{abstract}
\end{frontmatter}
\section{Introduction}
\label{sec:intro}
Social media has become an indispensable part of our daily life. People use social media to consume content that piques their interests. A plethora of content is available online in the form of texts, photographs, videos, podcasts, and so on.  It is claimed that people recall just 10\% of a message while reading, compared to 95\% while viewing a video\footnote{ https://www.forbes.com/sites/yec/2017/07/13/how-to-incorporate-videointo-your-social-media-strategy/}. Since human attention spans are getting shorter, micro-videos have emerged as a new and popular form of user-generated content. Due to the prevalence of cameras and smartphones, many people capture events of their daily lives and post them on various micro-video sharing services such as TikTok\footnote{\label{f1}https://www.tiktok.com} and Instagram\footnote{\label{f2}https://www.instagram.com}. Micro-videos are short and bite-sized video clips. These videos have large audience segments and are gaining traction in a way different from standard videos. 
TikTok has garnered 689 million monthly active users, and more than one million videos are viewed daily~\citep{basch2021videos}. Owing to the huge volume of micro-videos sprouting in micro-video sharing platforms day by day, managing the massive amount of micro-video data, including search and categorization has become imperative.

Hashtags i.e., keywords prefixed with a hash (\#) sign serve as organizational tools for managing the burgeoning repository of micro-videos disseminated across social networks. These hashtags augment the user experience by substantially ameliorating search functionality, enabling users to pinpoint micro-videos pertaining to specific subject matters. The utility of hashtags extends across a myriad of applications, encompassing query expansion, sentiment analysis, information retrieval, content discovery, and navigation. However, a noteworthy proportion of social media users refrain from affixing hashtags to their micro-videos, often owing to either a lack of awareness or an absence of concerted effort. Statistical data underscores the prevalence of this trend, revealing that more than 33 million hashtag-devoid micro-videos are posted daily on Instagram~\citep{wei2019personalized}. Moreover, an analysis of a dataset compiled by 
\citep{cao2020hashtag} indicates that nearly 65\% of the 5,84,876 micro-videos examined lack hashtags. In light of the escalating information deluge and the imperative to expedite video retrieval for users, developing a hashtag recommendation system tailored for micro-videos has become an exigent necessity. The primary objective of an automated hashtag recommendation system is to analyze the users' uploaded micro-videos and suggest pertinent hashtags.

Existing approaches to recommend hashtags for micro-videos can be classified into content-based~\citep{cao2020hashtag, djenouri2022deep, yang2020sentiment,yu2023generating} and personalized~\citep{li2019long,liu2020user,wei2019personalized} methods. Content-based approaches predominantly rely on intrinsic content characteristics while overlooking user preferences. Personalized systems harness user’s history, profile data, and individual user preferences to provide tailored hashtag recommendations. Given that personalized approaches require historical interaction data of users, these approaches are susceptible to cold-start user problems. Consequently, these approaches struggle to yield hashtag recommendations for cold-start users devoid of interaction history. In GCN-PHR~\citep{wei2019personalized}, authors model the user’s historical interactions with micro-videos and accompanying hashtags. However, this method exclusively centers on content and personalization while ignoring the interests of other similar users. The techniques combining content and personalization aspects emphasize the content and the impact of users’ preferences but largely overlook user interaction modeling, which can effectively capture valuable insights into user behavior within social networks. 

Modeling user interactions helps identify users with similar interests and behaviors.  This information is crucial for recommending hashtags that are relevant to the content and resonate with users' preferences and behavior patterns. This increases the likelihood that users will find the recommended hashtags meaningful and engaging.  Modeling user interactions is part of collaborative filtering, facilitating a profound comprehension of user behaviors. This comprehension, in turn, contributes to the refinement of personalized recommendations by considering how users with similar hashtag usage patterns engage with content. This added layer of personalization ensures that users are presented with hashtag suggestions that exhibit a close alignment with their idiosyncratic preferences, behavioral tendencies, and content consumption behavior. Hashtag recommendations aim to increase user engagement with micro-videos and the platform. Recommendations based on user interactions are more likely to resonate with users, encouraging them to explore and engage with content that aligns with their interests. This, in turn, leads to improved user engagement and satisfaction. New users who lack a history of interactions on a platform face challenges in receiving relevant recommendations, commonly known as the cold-start user problem. Modeling interactions among users can help alleviate the cold-start user problem by drawing insights from the behavior of users with established interaction histories. This enables the system to provide meaningful hashtag recommendations even to users who are new to the platform. Since social networks play a significant role in influencing user behavior, modeling user interactions can capture the effects of user communities and network dynamics. This helps in recommending popular hashtags within specific user groups or communities, further enhancing the relevance and engagement of the recommendations. These factors motivate us to develop a hybrid model offering complementary recommendations by leveraging content relevance, individual user preferences, and user interactions with fellow users. 

\begin{figure}[!h]
\centering
	\includegraphics[width=10cm, height=20cm,keepaspectratio]{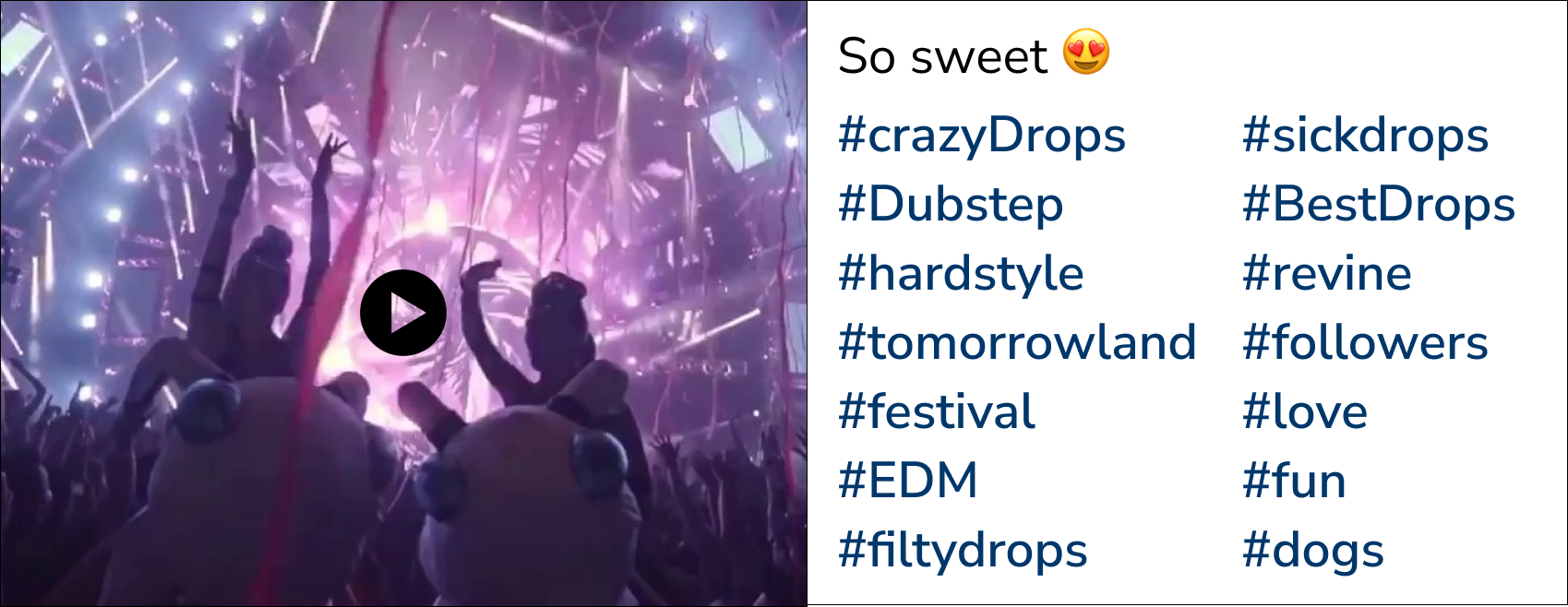}
	\caption{Example of a Micro-video from Vine with Corresponding Hashtags}
	\label{fig:eg}
\end{figure}

Furthermore, micro-videos contain a wealth of multimedia content embedded in textual, visual, and acoustic modalities. Each unimodal representation of the micro-video is equally significant in conveying the information contained in the micro-video. 
However, existing methods~\citep{cao2020hashtag,wei2019personalized,yang2020sentiment} involve the extraction of visual, acoustic, and textual features from micro-videos directly followed by concatenation to form an overarching micro-video representation. While this approach has been instrumental in the field, it has some limitations. Concatenating features from diverse modalities to create a unified representation for a micro-video inadvertently leads to the dilution of modality-specific information. Each modality, whether visual, acoustic, or textual, possesses unique characteristics and nuances that contribute to a comprehensive understanding of micro-video content. Blending them into a singular feature vector risks losing these distinctive traits, which are vital for analyzing the content to recommend appropriate hashtags. Existing approaches treat all modalities equally and do not account for their unique characteristics.
Let us illustrate this with an example micro-video of a live music performance as shown in~\autoref{fig:eg}.
This micro-video comprises visual, acoustic, and textual elements.
Visual features capture perceptible details such as performers' expressions, instrument playing, and audience reactions. Acoustic features encapsulate the rhythm, melody, and mood of the music. Textual features contain information about the festival and the crowd's enthusiasm. 
The associated hashtags are relevant since they describe the different aspects of the micro-video, including the music genre, the setting, the people, and the overall atmosphere. The acoustic hashtags i.e., \#dubtsep, \#hardstyle, and \#EDM relate to the type of music being played in the video, which is dubstep, and hardstyle. These are both electronic dance music genres that are known for their heavy bass and energetic beats. The visual hashtags i.e., \#festival, \#followers relate to the micro-video's setting, which is a music festival. This is evident from the large crowd, the stage, and the lights. The hashtag \#tomorrowland suggests that the micro-video may have been taken at the Tomorrowland music festival, which is known for its large crowds and elaborate stage productions. The textual hashtags i.e., \#love and \#fun are more general, but they are still relevant to the video because they describe the overall atmosphere, which is one of excitement and fun. The text ``So sweet" in the image suggests that the micro-video may be a positive or heartwarming video. This is consistent with the hashtags \#love and \#festival. In a concatenation-based approach, all these features are combined into a single vector. When these diverse features are concatenated, they blend together, potentially causing a loss of modality-specific nuances and difficulty in interpreting features. The fine-grained nuances of each modality may be overshadowed by the sheer volume of data from other modalities. For instance, the textual features i.e., subtleties of the crowd's reactions might be drowned out by the visual and acoustic elements.  The concatenated representation makes it challenging to interpret which features are contributing most to the final result. It is unclear whether visual, acoustic, or textual aspects are driving the recommendation.
In contrast, the approach of representing each modality as a separate node in a graph beside the user and modeling modality-to-modality and user-to-modality interactions, in order to update node embeddings, offers a solution to information dilution. By treating each modality as a separate node, the unique characteristics of visual, acoustic, and textual data are preserved. This makes it easier to interpret the contribution of each modality to the final micro-video representation and provides insights into whether visual, textual or acoustic cues play a more dominant role in recommending specific hashtags for micro-videos. In essence, the use of modality-specific nodes helps to mitigate information dilution by allowing each modality to maintain its individuality. This, in turn, leads to more informative, nuanced, and context-aware micro-video representations, ultimately benefiting the task of hashtag recommendation.

In this study, we propose a hybrid filtering graph-based deep neural network for MIcro-video haSHtag recommendatiON, i.e., MISHON, that can automatically recommend hashtags to unannotated micro-videos. The suggested methodology takes a three-pronged approach, i.e., content, personalized, and collaborative filtering. MISHON provides an advantage over content-based filtering by modeling user preferences besides user-user interactions. Additionally, it tackles one of the drawbacks of personalized systems and collaborative filtering by suggesting hashtags for micro-videos created by cold-start users. MISHON considers modality-specific information on micro-videos, users' individual preferences, and the tagging behavior of like-minded users to recommend suitable hashtags. 
The modality-specific representations of micro-video and user representations are refined by leveraging the inductive representation learning capability of graphs. The overall micro-video representation thus derived from these refined representations is utilized to recommend pertinent hashtags for micro-videos. 
The experiments conducted on three real-world datasets show the encouraging performance of our proposed framework. Our proposed model can recommend relevant hashtags and has a significant gain in performance. \\

We present the notable contributions of our work below.
\begin{itemize}
\item We propose a hybrid filtering-based system to recommend hashtags for micro-videos by bridging the gap inherent in content-based and collaborative filtering-based techniques. To this end, we leverage rich information embedded in the micro-video content and interactions among similar users. 
\item We devise a novel way of modeling collaborative filtering by leveraging hashtag usage styles to find the most similar users. 
\item To overcome the pervasive cold-start user problem in collaborative filtering, we propose a social influence technique and recommend relevant hashtags for micro-videos posted by cold-start users. 
\item We design a novel interaction graph to capture the modality-to-modality and user-to-modality interactions besides user-to-user interactions. We subsequently refine representations of the micro-video modalities and the corresponding user. 
\item Extensive experiments performed on three real-world datasets manifest the competitive advantage of the proposed framework against the state-of-the-art approaches.
\end{itemize}

The remainder of the paper is organized as follows.~\autoref{sec:rw} goes over the related works.~\autoref{sec:pd} defines our problem setting and formulation. In~\autoref{sec:methodology}, we present our technique.~\autoref{sec:expandresults} then shows the experimental evaluations followed by discussion in~\autoref{sec:discussion}. Finally, in~\autoref{sec:conclusions}, we conclude our work.

\section{Related Work}
\label{sec:rw}
This section covers the substantial volume of research work done in the domain of hashtag recommendation. Current hashtag recommendation methodologies can be categorized into three distinct classes, specifically content-based, personalized, and collaborative filtering.

\subsection{Content-based Hashtag Recommendation}
Content-based hashtag recommendation strategies endeavor to explicitly represent the substantive attributes of an item in order to propose hashtags for that item. This approach has been subjected to rigorous examination across a diverse spectrum of content types, including textual data, visual media, and micro-videos.

\subsubsection{Text-based Hashtag Recommendation}
\citep{kumar2021hashtag} suggested hashtags using word embeddings and additional knowledge from external sources i.e., Wikipedia and the Web to close the gap between tweet content and associated hashtags. The authors retrieved semantically related keywords from word embeddings using word2vec besides using topical, lexical, semantic features of tweets and user influence. 
\citep{zhu2019learning} turned to the adoption of Tree-LSTM, a specialized architecture, with the intention of augmenting the representation of textual content and, consequently, refining the quality of hashtag recommendations. 
\citep{zhang2023twhin} devised a bipartite graph that interlinked tweets and users, serving as the foundation for the exploration of socially akin tweets and the prognostication of hashtags for multilingual content. Meanwhile, 
\citep{bansal2024multilingual} proposed a polyglot graph-based deep neural network, tailored to the intricate task of suggesting hashtags for tweets composed in low-resource Indic languages. 
\citep{chakrabarti2023hashtag} introduced an approach to recommend contextually appropriate hashtags with the aim of augmenting social visibility. Their method leverages the trend dynamics of hashtags by integrating post keywords in conjunction with user and post popularity as determinants for recommending hashtags.

\subsubsection{Image-based Hashtag Recommendation}
\citep{chen2021tagnet} created an image similarity graph to illustrate the relationship between posts assuming visually comparable images use similar hashtags. The Triplet Attention module captures the influence of multiple modalities to derive node features. The aggregated graph convolution component learns the attended features and spreads information among vertices to suggest suitable hashtags. 
\citep{lakshmi2023hybrid} introduced a hybrid recommendation framework for social image tagging, which is founded on a classifier-driven ontology approach. The model leverages the XGBoost classifier and Gated Recurrent Units (GRUs)~\citep{cho2014learning} for the classification of instances. This comprehensive framework encompasses elements of semantic understanding, knowledge consolidation, and an ontology-driven approach. 
\citep{jafari2023popular} harness transformer models to produce hashtags for photographs in a manner that aligns with the prevailing popularity of hashtags. In the initial phase, multilabel classification algorithms are deployed to identify constituent elements within images and extract relevant image attributes. This initial phase serves as the foundation for generating hashtags. Subsequently, in the second stage, hashtags are produced using transformer-based text generation models, with the level of hashtag popularity being a pivotal factor in the generation process. 

The aforementioned systems utilize information from either texts or images for recommending tags. Relatively fewer studies have been conducted to recommend hashtags for micro-videos. Hashtag recommendation for micro-videos is more challenging than the text-only or image-only methods. In this paper, we consider the multimodality of micro-videos to harness the complete spectrum of information encapsulated within modalities constituting the micro-videos to improve the rationality and accuracy of hashtag recommendations. 

\subsubsection{Video-based Hashtag Recommendation}
\citep{mehta2021open} devised an open-domain trending hashtag recommendation system for micro-videos. Micro-videos and hashtags are used as nodes in a heterogeneous graph. The edges between hashtag nodes are created based on their co-occurrences. The node state update scheme in GCN is employed to generate the final micro-video embedding which is then used to advocate hashtags. 
\citep{yu2023generating} formulated the task of hashtag recommendation for micro-videos as a generation task that better represents how hashtags are created naturally. The authors propose a Guided Generative Model (GGM) that generates hashtags from multimodal inputs and guided signals from a VLM-based Hashtag Retriever. 
\citep{cao2020hashtag} developed a Neural Network-based framework for micro-video hashtag recommendation. The authors first obtain feature representations of micro-video modalities and integrate them via a multi-view representation learning framework. The regularized projections and hashtag embeddings are fed to a customized neural collaborative filtering framework that yields hashtags to be recommended for each micro-video. 
\citep{yang2020sentiment} developed a model that comprises content and sentiment common space learning branches. The sentiment features, content features, and semantic embeddings of hashtags are integrated using a weighted concatenation. 

The content-based algorithms ignore the user information, and recommend hashtags for the item by solely considering the information related to the item’s content. These systems may not take into account the individual preferences, habits, or behaviors of users. Thus, they might not provide personalized recommendations that align with a user's specific interests. Moreover, micro-videos are frequently shared and consumed in social contexts. Content-based models do not typically account for the social interactions and relationships between users, which are vital for understanding the popularity and relevance of hashtags. When dealing with new micro-videos, content-based models may struggle to provide meaningful hashtag recommendations due to the limited available data for analysis.

\subsection{Personalized Hashtag Recommendation}
~\citep{liu2020user} introduced an innovative framework for micro-video hashtag recommendation that considers the content of the micro-videos and the distinct user preferences associated with them. The framework assimilates user information, encompassing user-profiles and historical hashtag usage, into the process of learning micro-video embeddings. This integrated approach perceives user-related data as an input query and adeptly cultivates attention mechanisms at both image-level and video level to enrich the representations of micro-videos. 
\citep{bansal2022hybrid} introduced a novel multimodal personalized hashtag recommendation technique. Their approach integrates pertinent data derived from multiple modalities within social media posts and encapsulates user preferences in the pursuit of suggesting plausible hashtags. The authors introduce a hybrid deep neural network that capitalizes on hashtags through a fusion of multilabel classification and sequence generation methodologies, thus facilitating the recommendation of a curated list of candidate hashtags for social media posts. 
\citep{li2019long} introduced a novel multiview representation interactive embedding approach and proliferated data using graphs to advocate hashtags for micro-videos. The proposed system simultaneously considers sequential feature learning, micro-video-user-hashtag interaction, and hashtag correlations. The authors created a hashtag graph using external knowledge to address the long-tail distribution of hashtags. Subsequently, an interactive embedding network is employed to predict interactions of hashtags with micro-videos. To recommend personalized hashtags, 
\citep{wei2019personalized} used Graph Convolutional Network (GCN)~\citep{hamilton2017inductive} to simulate intricate relationships among users, micro-videos, and hashtags. The authors construct a heterogeneous graph in which nodes are micro-videos, users, and hashtags, edges are based on their direct relationships, and a message-passing strategy is adopted to modify node representations. The proposed model further learns micro-video features and hashtag embeddings conditioned on obtained user preference and hashtag representations to suggest personalized hashtags for micro-videos. 

Personalized models can overfit a user's past behavior, potentially missing out on broader interests. These systems may struggle to provide meaningful recommendations for new or cold-start users due to limited interaction data. We model the interaction among users and leverage collective user behavior to fill this data gap. Interaction modeling among users helps overcome these limitations by considering the collective behavior of users, making recommendations more robust, diverse, and adaptive. Interaction modeling can provide more generalized recommendations that are less susceptible to overfitting. Moreover, our proposed model’s variant tackles the cold-start problem inherent in personalized hashtag recommenders. 

\subsection{Collaborative Filtering-based Hashtag Recommendation}
\citep{torres2020hashtags} 
employed Collaborative Filtering (CF) for suggesting hashtags. The authors devised the memory-driven KNN algorithm where both users and hashtags are used to predict the missing values in the user-item matrix. To unite users who share the same interests, 
\citep{alsini2020utilizing} summarized community discovery techniques. The study looked into the effects of four relationships namely, hashtag usage, subjects, followers, and mentions on the efficacy of hashtag suggestion across communities. 
\citep{yong2022implicit} presented a user-centric tag recommendation algorithm based on CF. The algorithm exploits the LDA topic model to derive implicit tags from item text and computes feature weights for item-tag associations. Through the integration of the item-tag matrix and the user scoring matrix, the authors derive the user's preference matrix for tags. To solve the dearth of user-item interactions in e-commerce recommender systems, 
\citep{zheng2022multiview} presented the Multiview Graph CF network that harnesses the combined power of homogeneous and heterogeneous signals obtained from attribute and neighbor perspectives. The authors use co-occurrence characteristics of varied attribute values and cooperative preferences among several neighbors to obtain useful embedding representations of nodes. 
\citep{zhou2022hybrid} devised hybrid CF for generating dynamic service suggestions in a mobile cloud environment while considering user preferences. The first step was to extract user preferences from quality of service (QoS) data and use them as a factor for grouping similar customers. The top-k approach was then used to find the target users' and products' related neighbors, making it easier to anticipate users' QoS values.
CF struggles when faced with new items or users that have limited or no data available referred to as cold-start problem~\citep{he2023usbe,ur2023caml}. 
\citep{belem2019exploiting} leveraged the synergistic effects of syntactic and neighborhood-based characteristics to bolster and refine tag recommendation techniques, specifically in scenarios characterized by a cold-start. The researchers extracted innovative tag quality attributes from the syntactic patterns inherent in the textual content linked to Web 2.0 objects, with the goal of identifying and offering pertinent tag suggestions.
\citep{duan2023multi} devised a collaborative attention network with many features merged to produce sequential recommendations. The framework captures the temporal and spatial properties of nodes to analyze their dynamic patterns. To strengthen the semantic links between nodes and reduce noise from the structural neighborhood, semantic-enriched contrastive learning is also used.

Collaborative filtering, by design, overlooks individual user preferences and context. Interaction modeling allows for more personalized recommendations by considering hashtag usage and user interactions, making recommendations better aligned with users' interests. 
By modeling interactions among users based on hashtag usage, similar modality interaction, and user interactions with the modality of their posted micro-videos, the limitations of collaborative filtering in the context of hashtag recommendation for micro-videos can be effectively addressed. This approach enables more robust, diverse, and personalized recommendations while mitigating issues related to data sparsity and the cold-start user problem. We devise a content and social influence-based solution to yield hashtag suggestions for new micro-videos posted by cold-start users. Our hybrid hashtag recommender for micro-videos aims to strike a balance between content, users’ individual preferences, and the hashtag usage behavior of like-minded users.

Our work is complementary to the above-mentioned works as we are putting forth a hybrid hashtag recommender system powered by deep learning and GNNs that utilizes both content and collaborative filtering techniques to recommend relevant hashtags for micro-videos. We construct a novel interaction graph and employ GNN to model the interaction among users and constituent modalities of micro-videos to yield pertinent hashtags. Furthermore, we propose a social influence and content-based technique to alleviate the cold-start user issue that plagues collaborative filtering-based recommender systems.
\section{Problem Definition}
\label{sec:pd}
This paper proposes a hybrid filtering-based hashtag recommender for micro-videos. Existing works ignore the importance of modeling relatedness among like-minded users. Therefore, the proposed model, MISHON has the following objectives: 
(1) to recommend relevant hashtags for micro-videos while considering content and user relatedness.
(2) to address the cold-start user issue that prevails in collaborative filtering
(3) to model interactions among micro-video modalities and users and better capture the semantic information among nodes.


Consider a social media dataset $D$ with the following attributes: a micro-video set $M= \{mv_i\}^{\lvert M \rvert}_{i=1}$, a hashtag set $H=\{hg_j\}^{\lvert H \rvert}_{j=1}$, and a user set $U=\{u_k\}^{\lvert U \rvert}_{k=1}$. {\it Given a micro-video post $(mv_i)$ uploaded by a user $(u_k)$, we seek to automatically recommend a collection of hashtags $R = \{rh_r\}^{\lvert R \rvert}_{r=1}$ that is credible and corresponds to the set of ground-truth hashtags $G=\{gh_g\}^{\lvert G \rvert}_{g=1}$.}

Here, ${\lvert . \rvert}$ stands for the cardinality of a set. The variables $i$, $j$, and $k$ serve as indices for the micro-video post, hashtag, and user correspondingly, while $r$ and $g$ function as indices for the recommended and ground-truth hashtags.
Hashtag recommendation for an existing user and the cold-start user is defined as follows. 

\vspace{0.05in}
\noindent {\bf Problem 1} {\it(Hashtag Recommendation for an Existing User)} {\it Given a new micro-video post $(mv_{L+1})$, where $mv_{L+1} \in M$ created by a user $(u_k)$ who posted $L$ micro-videos in the past, we intend to suggest appropriate hashtags for the new micro-video post $(mv_{L+1})$ automatically}.

Hashtag recommendation requires us to suggest suitable hashtags for a micro-video. Hashtags function as abstract labels that represent the information contained in each modality. Here, we attempt to recommend a plausible set of hashtags for a new micro-video posted by an existing user. 
We take the content-based information embedded in constituent modalities of the micro-video besides user-to-modality and user-to-user interactions.

\vspace{0.05in}
\noindent {\bf Problem 2} {\it(Hashtag Recommendation for a Cold-start User)} {\it Given a target micro-video post $(mv_i)$ created by a cold-start user $(u_{\lvert U \rvert+1})$, our aim is to automatically recommend a relevant set of hashtags for the micro-video posted by that user.} 

A cold-start user problem occurs when a new user joins the platform. In this situation, making recommendations is exceedingly challenging since there is no information regarding the user's previously created micro-videos and tagging habits. Collaborative filtering cannot make meaningful hashtag recommendations because these methods look for specific users who have similar interests based on past behavior. Our study attempts to overcome the constraints of collaborative filtering by devising a social influence and content-based technique. Here, we solve the cold-start problem inherent in hashtag recommendation systems by recommending appropriate hashtags for a new micro-video post $(mv_i)$ created by a cold-start user $(u_{\lvert U \rvert+1})$.


\vspace{0.05in}
\noindent {\bf Problem 3} {\it(Feature Refinement)}
{\it Given the fine-grained modality-specific representation ($mv_{initial}$) and initial user representation ($u_{initial}$) of the target user ($u_{k}$), our objective is to train a function $g$(.) that models various interactions among users and micro-video modalities such that:}

\begin{equation}
mv_{final}, u_{final} = g(mv_{initial}, u_{initial})
\label{eq:eq3}
\end{equation}
Here, $mv_{final}$ and $u_{final}$ represent the co-influential feature vector of micro-video and user respectively. The main challenge of working with heterogeneous data is modeling the interaction of multiple data types and then combining them into one uniform embedding. Micro-videos are heterogeneous sources of data comprising three modalities. Users assign hashtags according to their topics of interest in every modality as well as shared interests with other users. Therefore, the modality-specific and user representations play key roles in obtaining the overall micro-video representation. Here, we attempt to learn a function that can model interactions among constituent modalities of micro-videos and users to refine the overall representation of the micro-video and recommend suitable hashtags.

\begin{figure}[!h]
\centering
	\includegraphics[width=15cm, height=20cm, keepaspectratio]{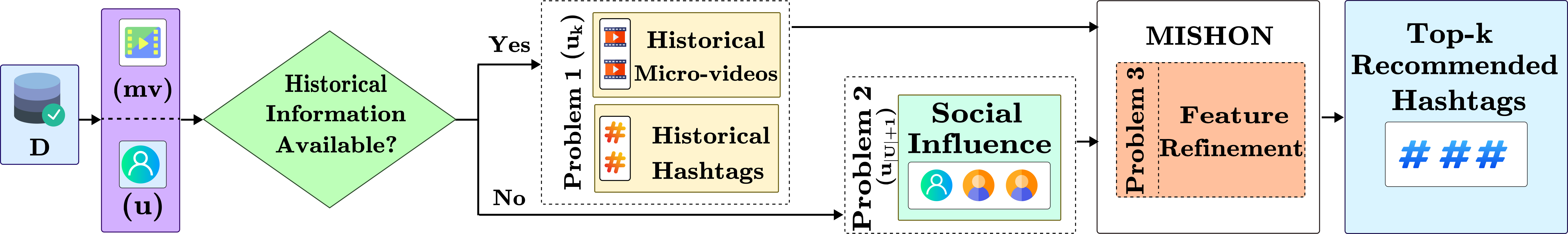}
	\caption{Visual Representation of Problem Definition}
	\label{fig:pd}
\end{figure}

The above-mentioned problems have been pictorially depicted in~\autoref{fig:pd}.
\section{Methodology}
\label{sec:methodology}

\begin{figure}[!ht]
	\includegraphics[width=\textwidth]{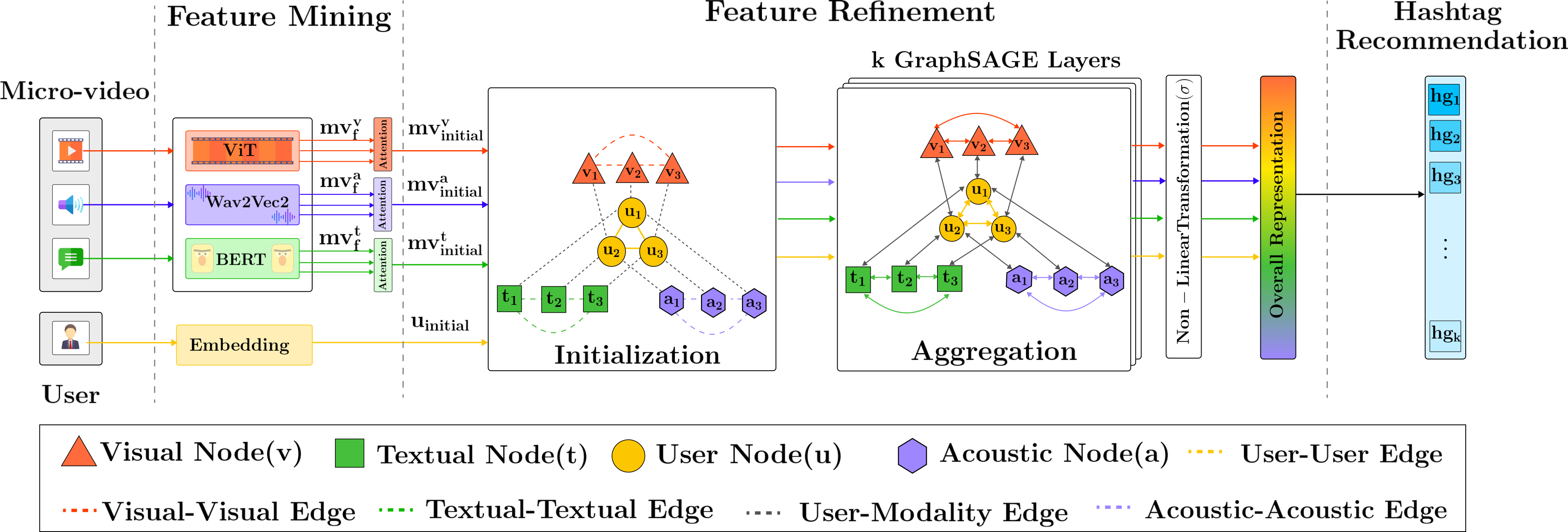}
	\caption{Overall Architecture of MISHON}
	\label{fig:sysarch}
\end{figure}

In this section, we elucidate our proposed approach.
We propose an integrated model that incorporates micro-videos and users to tackle the task of micro-video hashtag recommendation. We begin with a high-level overview of our system as shown in~\autoref{fig:sysarch} before delving into its components. The input to our system is a micro-video post and the corresponding user who created that micro-video post. The micro-video post is divided into its constituent modalities. The modality-specific features of the micro-video are extracted through respective feature extractors. We then employ an attention mechanism on the modality-specific features to find information that is most apt to recommend hashtags. 
We enhance micro-video representation by constructing an interaction graph that comprises four types of nodes and seven types of edges to capture modality-to-modality, user-to-user, and user-to-modality interactions.
The initial node embeddings are updated based on information propagation and neighborhood aggregation. The overall representation thus obtained, is fed into the hashtag recommendation module as input. After taking into account the likelihood of each hashtag, the hashtag suggestion module produces a sorted list of ``top-K" hashtags. As demonstrated in~\autoref{fig:sysarch}, our proposed framework comprises three components: (a) feature mining; (b) feature refinement; and (c) hashtag recommendation. Below, we go through each component in further detail.
\subsection{Feature Mining}
\label{ssec:fm}
This section describes the feature mining module that is made up of two submodules: (a) feature extraction and (b) attention modeling. We first extract features of modalities constituting the micro-video. We endeavor to enrich modality-specific representations followed by an attention mechanism to filter out noisy information. We describe the details of each submodule below.
\subsubsection{Feature Extraction}
\label{sssec:fe}
In this, we elaborate on details of feature extraction from modalities constituting the micro-video. We first segment every micro-video into visual, acoustic, and textual modalities denoted by $mv_i^v$, $mv_i^a$, and $mv_i^t$ respectively. We then retrieve features corresponding to each modality.

\textit{Visual Feature Extraction:}
\label{p:vfe}
Micro-videos are only a few seconds long.
Due to the concise nature of micro-videos, a limited number of keyframes can effectively encapsulate the entirety of the visual content. To extract the micro-video frames, we use  FFmpeg\footnote{https://www.ffmpeg.org/} and extract $12$ frames for each micro-video at equally spaced intervals. We employ Vision Transformer~\citep{dosovitskiy2021an} to derive visual attributes for every frame of the micro-video. Vision Transformer, abbreviated as ViT, is a variant of the language-based transformer model that takes an image as an input, uses the image structure to learn meaningful embeddings, and performs image classification. We employ the basic Vision Transformer architecture with a $16 \times 16$ input patch size for the frame feature extraction of every micro-video. We rescale every frame to meet the model's requirements for input size. ViT creates a grid of square patches to split the frame. The channels of all pixels in a patch are concatenated, and the patch is then linearly projected to the chosen input dimension, flattening each patch into a single vector. Since transformers do not take into account the input element's structure, therefore we give each patch learnable position embeddings so the model can pick up on the image's structure. There are 12 attention modules in total. It is important to note that ViT does not contain any convolutional layers. 

Given a sequence of frames representing the visual modality $mv_i^v$ of the micro-video $mv_i$, we employ the pretrained ViT to extract frame features. We regard the penultimate layer of ViT to obtain visual features.
\begin{equation}
    mv_f^v=ViT(mv_i^v)
    \label{eq:eqfe1}
\end{equation}
Here, $mv_f^v \in \mathbb{R}^{N_f \times D}$ represents the resultant visual feature matrix, where $N_f=12$ stands for the number of frames and $D=768$ for the concealed size of each frame.
\textit{Acoustic Feature Extraction:}
The acoustic modality, as a crucial supplement to the visual modality, is especially effective when the visual content is too diversified or conveys inadequate information. To capture the acoustic characteristics, we separate the audio channel from the video and subsequently divide it into equidistant segments of uniform duration using FFmpeg. Following that, we employ wav2vec2.0~\citep{baevski2020wav2vec} to extract features for each audio clip. 
Wav2vec2.0, a self-supervised speech representation model, aims to capture essential characteristics of unprocessed audio files by harnessing the strength of transformers and contrastive learning.  This method is comparable to the masked language modeling used in Bidirectional Encoder Representation from Transformer, abbreviated as BERT~\citep{kenton2019bert}. Wav2vec2.0 can obtain high-level contextual representation and learn basic units for less labeled data. There are two stages to the wav2vec2.0 training procedure: in the first stage, the model is trained on hundreds of unlabelled data, and in the second stage, it is fine-tuned on a small dataset for certain tasks. The wav2vec2.0 model has the following components: convolutional layers that turn the raw waveform input into latent representation $(Z)$; transformer layers that produce contextualized representation $(C)$ and linear projection produces the output $(Y)$. Wav2vec2.0 uses a multilayer Convolutional Neural Network (CNN) to extract latent audio representations of 25 ms each from raw audio data. For feature extraction and selection, the representations are contained in a quantizer and a transformer. Gumbel and K-means are used to quantify data. Every 20 ms, the wav2vec2.0 toolkit extracts a 768-dimensional feature vector from the voice stream for a certain encoder layer. Each layer produces a new representation, which can vary in suitability for a job than a preceding or succeeding layer.
We convert the raw audio from .mp3 to .wav format to satisfy the model's input requirements, and the sampling rate is preserved at 16,000 Hz. We employ the base version of the wav2vec2.0 model that is pre-trained on 960 hours of unlabelled speech from the LibriSpeech~\citep{panayotov2015librispeech} corpus. 

Given the acoustic modality of the micro-video denoted by $mv_i^a$, we apply wav2vec2.0 to extract acoustic features.
\begin{equation}
    mv_f^a=Wav2vec2.0(mv_i^a)
    \label{eq:fe2}
\end{equation}
Here, $mv_f^a$ represents the resultant acoustic feature matrix where $mv_f^a \in \mathbb{R}^{299 \times D}$ and $D=768$ represents the embedding size. We chose to extract acoustic features from the penultimate layer. We obtain a 768-dimensional feature vector for the entire audio segment of a given micro-video. 

\textit{Textual Feature Extraction:}
Textual descriptions play a pivotal role in providing information about the micro-video post from a different perspective. Textual modality has established its importance for hashtag recommendation as demonstrated by previous works~\citep{jafari2023unsupervised,cantini2021learning, djenouri2022toward, kumar2021hashtag}. The text encoder generates final text representations from the natural language sentence i.e., a textual description of the micro-video post. For the text encoder, we employ a Transformer-based model i.e., BERT.  
For the textual modality of micro-video denoted by $mv_i^t$ which comprises a word sequence, we add two tokens: class (CLS) and separator (SEP) that mark the start and end of the input text, respectively. We generate the corresponding set of tokens $T$ using the BERT tokenizer.

\begin{equation}
T = BERT\_Tokenizer(mv_i^t)
\end{equation}
We set a 30-token cap on the length of the token sequence $S$. For textual descriptions less than $S$, we apply padding, otherwise, we perform truncation to make all textual descriptions of uniform size. Finally, we create token-based embeddings using BERT, as depicted in~\autoref{eq:tfb}.

\begin{equation}
mv_f^t = BERT(T)
\label{eq:tfb}
\end{equation}
The final textual feature matrix 
$mv_f^t \in \mathbb{R}^{S \times D}$, where $S =30$ denotes the maximum length of the associated text for the micro-video post, and $D =768$ denotes the embedding size.
\subsubsection{Attention Modeling}\label{mishon-attention}
Hashtags are typically used to emphasize significant information in micro-videos. As a result, eliminating noisy information and determining the importance of each unit constituting the modality-specific representation is critical in the hashtag suggestion task. Due to the effectiveness of attention mechanism~\citep{vaswani2017attention}, we apply it individually on every modality as given in~\autoref{eq:eqatm}.
\begin{equation}
    mv^m_{initial}=Attention(mv^m_{f})
    \label{eq:eqatm}
\end{equation}
The enriched modality-specific feature vector in this instance, $mv^m_{initial}$, was acquired via an attention method.  The modality-specific embedding can be thought of as a sequence of feature vectors as shown in ~\autoref{eq:eq_t}. 
\begin{equation}
   mv^m_{f}=\{mv^m_x\}_{x=1}^X  
   \label{eq:eq_t}
\end{equation}
where $X$ denotes the number of units in every modality. We feed each unit $mv^m_x$ to a Multi-Layer Perceptron (MLP) to get $h^m_x$ as a hidden representation of $mv^m_x$, as shown in~\autoref{eq:eqat1}.
By assigning an attention weight to every unit in each modality, we explicitly represent its varied relevance. To create an enriched representation of the constituent modality, we extract key units in each medium and combine the resultant unit representations.
\begin{equation}
    h^m_x = tanh(Wmv^m_x + b_w)
    \label{eq:eqat1}
\end{equation}
Here, $h^m_x$ is the concealed representation of $mv_x^m$. We compute each unit's relevance ($\alpha^x$) as shown in~\autoref{eq:eq9}.
\begin{equation}
     \alpha^x =  softmax\left(\left(h^m_x \right)^T u_w\right)
     \label{eq:eq9}
\end{equation}
In this case, $\alpha^x$ symbolizes the importance of a unit while $u_w$ denotes the context vector. To obtain the standardized coefficient ($\alpha^x$), we initially calculate the resemblance between $h^m_x$ and the contextual vector ($u_w$). We subsequently subject the result to a softmax function for normalization. The enriched modality-specific feature vector is then calculated, as shown in~\autoref{eq:eq10}.
\begin{equation}
mv^m_{initial} =\sum_{x=1}^{X}\alpha^x{h^m_x}   
\label{eq:eq10}
\end{equation}
Here, $mv^m_{initial}$ denotes the enriched modality-specific feature vector which is obtained by aggregating annotations using the coefficients $\alpha^x$. Further, the user-to-modality and user-to-user interactions can help obtain a better micro-video representation. To facilitate the learning of these interactions, we employ a graph neural network as discussed in~\autoref{ssec:fr} in order to refine modality-specific and user representations.
\subsection{Feature Refinement}
\label{ssec:fr}
In this section, we elaborate on the feature refinement module. It consists of three steps namely:
(1) graph construction; (2) information propagation and neighborhood aggregation; and (3) micro-video representation. We discuss these steps below.
\subsubsection{Graph Construction}
We construct an undirected graph $G = (N, E)$, where $N$ and $E$ denote the collection of vertices and edges, respectively. The total number of nodes in the graph is $I$ where $I=3M+U$ and $E \subset I \times I$ is a set of relationships containing their interdependencies.
We further elaborate on graph construction in the following sections. 

\textit{Node Settings:}
We construct a graph with four different kinds of nodes as shown below. 
\begin{equation}
 N=V \cup A \cup T \cup U
 \label{eq:eq4}
\end{equation}
Specifically, $N$ comprises four different types of entities: $V$, $A$, $T$, and $U$, where $V$, $A$, and $T$ represent the set of visual, acoustic, and textual modalities constituting the micro-videos contained in the micro-video set ($M$), and $U$ represents the set of corresponding users. 
The micro-video ($mv$) is represented by three nodes $v, a, t$ corresponding to three modalities, initialized with $mv^v_{initial}, mv^a_{initial}, mv^t_{initial}$ respectively. The feature vectors are derived as discussed in~\autoref{ssec:fm}. The user who created the micro-video is considered the fourth type of node in the graph. The user id is transformed into a fixed-size vector representation $u_{initial} \in \mathbb{R}^D$, which is randomly initialized and refined throughout the training.

\textit{Edge Generation:}
When two nodes $n_i$ and $n_j$ interact, an edge $e_{ij} = (n_i,n_j ) \in E$ is formed to link two nodes in the network. To exploit dependency amongst
different kinds of nodes, we consider homogeneous and heterogenous modeling among them. To differentiate them, we use distinct edge weighting strategies.

(i) Homogeneous Modeling: 
This modeling refers to edges connecting the same type of nodes. There are four types of interactions: video-to-video, audio-to-audio, text-to-text referred to as intramodality edges, and user-user edges. We create weighted edges to link two nodes in the graph and set a threshold to filter out low-weighted edges. We discuss edge construction criteria for modeling modality-to-modality and user-to-user interaction below. 
\begin{itemize}
\item Intramodality edges: We create edges to connect nodes from the same modality of different micro-videos. Given the modality-specific feature representations of two micro-videos denoted by $(mv^m_f)_i$ and $(mv^m_f)_j$, where $m=\{v, a, t\}$ is the modality indicator for micro-video $(mv)$, the edge weight $e(m_i, m_j)$ is assigned as shown in~\autoref{eq:ime}.
\begin{equation}
     sim(m_i,m_j) = cs\left(\left(mv^m_f\right)_i, \left(mv^m_f\right)_j\right) 
     \label{eq:ime}
\end{equation}
Here, $sim(m_i,m_j)$ refers to the similarity score between the same modality of different micro-videos, $cs((mv^m_f)_i, (mv^m_f)_j)$ is the cosine similarity value between the $m^{th}$ modality-specific feature representations of $i^{th}$ and $j^{th}$ micro-videos. We create intramodality edges as exhibited in~\autoref{eq:eq7}.
\begin{equation}
  e(m_i,m_j) =
    \begin{cases}
      sim(m_i,m_j),\text{ if } sim(m_i,m_j)  \geq \theta\ \\
      0,\text{ if } sim(m_i,m_j) < \theta
    \end{cases} 
    \label{eq:eq7}
\end{equation}
Here, $e(m_i,m_j)$ refers to the weight assigned to edges connecting the same modality of different micro-videos denoted by $m_i$ and $m_j$ and $\theta$ refers to the threshold. We use cosine similarity value to assign weight to the edge between two nodes of the same modality.  We assume that if two nodes have higher similarity, they contain rich information. To this end, we set threshold $\theta$ to 0.5 to filter out edges with low semantic similarity.
\item{User-user edges}:
Under user-to-user interaction, we first discuss the correlation of an existing user followed by a cold-start user with other users on that micro-video sharing platform.

\textbf{Existing User}: The user who has already posted micro-videos on the video-sharing platform is considered as an existing user. We model co-occurrence relationships among existing users based on common historical hashtags. The historical hashtag set of an existing user comprises all hashtags used by him/her in previously posted micro-videos. We take collaborative filtering into account which assumes that users who have had similar interests in the past will have similar interests in the future. Users with similar interests tend to assign similar hashtags to their micro-videos since hashtags reflect user preferences from different granularities. The tagging behavior of each user is hidden in user co-occurrence relationships. 
We compute the similarity among users as illustrated below.
\begin{equation}
sim(u_i,u_j) = |H_i \cap H_j|/|H_i \cup H_j|
\label{eq:us}
\end{equation}
Here, $sim(u_i,u_j)$ is the similarity score of two users, $\cap$, $\cup$, and $\lvert. \rvert$ denotes intersection, union operators, and set cardinality. $H_i$ and $H_j$ where $H_i \subset H$ and $H_j \subset H$ is the set of historical hashtags of user $u_i$ and user $u_j$ respectively. The numerator in~\autoref{eq:us} denotes the number of common hashtags of two different users for their uploaded micro-videos and the denominator denotes the total number of hashtags contained in the set of their historical hashtags. The interaction modeling between two users is carried out as depicted in~\autoref{eq:eq6}.
\begin{equation}
  e(u_i,u_j) =
    \begin{cases}
      sim(u_i,u_j),\text{ if } sim(u_i,u_j) \geq \gamma\ \\
      0, \text{ if } sim(u_i,u_j) < \gamma
    \end{cases} 
    \label{eq:eq6}
\end{equation}

Here, $e(u_i,u_j)$ denotes the weight assigned to the edge connecting two different users i.e., $u_i$ and $u_j$. We assign an edge weight to model the degree of relatedness of users. We also assign a threshold $\gamma$ to filter out edges with a low weight. Here, the threshold $\gamma$ is set to $0.5$.

\textbf{Cold-start User}: Users who are new to the system and lack historical information are called cold-start users. The user’s historical interactions contain his interests based on which recommendations can be made. However, such interactions are often sparse, leading to cold-start user problems where a user has no historical posts and hashtags. Owing to the unavailability of user history on posted micro-videos and hashtags used, we employ a social influence technique to model the interaction of cold-start users with other users on that platform. People having high popularity are conceived as influential and more credible. The hashtagging patterns of influential users are mimicked by other users to garner social attention. We devise Algorithm~\autoref{alg1} to construct user-user edges and associate cold-start users with other users, assuming that cold-start users tend to utilize hashtags as used by the most popular users to increase their content's visibility and garner attention from other users on that platform. To determine a user's popularity, we compute the engagement rate (Line 2), which is the ratio of the total number of likes on the user's profile to his total number of followers. For a cold-start user, we set this value to 0. We then sort these users based on their engagement rate (Line 4), and only the top 10\% (Line 5, 6) are considered popular. Here, $argsort()$ returns the corresponding users with engagement rates sorted in descending order. Edges are constructed between popular users and cold-start user (Lines 8 to 13). The final set of obtained edges is denoted by ($E_{user}$) as shown in Line 14.

\end{itemize}
\begin{algorithm}[!ht]
\caption{Addressing Cold-Start User Problem}
\begin{tabular}{ll}
    \textit{Input:} & $U$: List of users\\
    & $\alpha=0.1$: Percentage of popular users\\
    & Metadata of user $u_i \in U$:\\
    & \hspace{0.5cm} Number of likes $(l_i)$\\
    & \hspace{0.5cm} Number of followers $(f_i)$\\
    \textit{Output:} & User-user edges $(E_{user})$\\
    \end{tabular}
\begin{algorithmic}[1]
    \FORALL{$ i=1$  to $\lvert U \rvert$}
        \STATE$Er_i=l_i/f_i$
     \ENDFOR
     \STATE$users\_sorted=argsort(Er)$
     \STATE$top\_p=\alpha* \lvert U \rvert$
     \STATE$popular\_users = users\_sorted[1 \hdots top\_p ]$
     \STATE${E_{user} = []}$
     \FORALL {$pu_i \in popular\_users$}
        \FORALL {$u_i \in U$}
             \STATE$E_{user}.append(pu_i, u_i)$
             \STATE$E_{user}.append(u_i, pu_i)$
         \ENDFOR
    \ENDFOR
    \STATE$\textbf{return} \; E_{user}$
\end{algorithmic}
\label{alg1} 
\end{algorithm}

(ii) Heterogeneous Modeling: This modeling refers to connecting different types of nodes. To interchange high-level semantic information among users and micro-video modalities, we model interactions among them. 
User-to-modality edges are drawn between users and constituent modalities of their uploaded micro-videos. An edge exists between user node $u_i$ and modality node $m_i$, where $m =\{v, a,t\}$ if the modality $m_i$ constituting the micro-video $mv_i$ was posted by user $u_i$. All edges are undirected with weights assigned to one for convenience. To explicitly model the user’s preference in modalities of his created micro-video, we construct the following edges: u-v, u-t, and u-a. 
We represent the interaction between the user and constituent modalities of his uploaded micro-video as $e(u_i,m_i)$ if $m_i \in mv_i$ and $mv_i$ is a micro-video created by user $u_i$. Here, $m$ is the modality indicator of micro-video $mv_i$.

\subsubsection{Information Propagation and Neighborhood Aggregation}
Deep Learning and graph techniques are
combined to handle recommendation-related problems. Raw data is transformed into structured graph form, after which an embedding process learns node embeddings of the graph. These updated embeddings are subsequently fed into a Deep Neural Network to provide recommendations. Essential characteristics of either node, edges, local node neighborhoods, or the entire graph can be encoded through meaningful embeddings. One of the most popular approaches to learning graph-structured data is GraphSAGE (SAmple and aggreGatE)~\citep{hamilton2017inductive}. The GraphSAGE framework is capable of creating informative node embeddings. Nodes in the same neighborhood should have similar embeddings, according to GraphSAGE's presumption. Based on a broad inductive framework, GraphSAGE updates node representations by considering the neighborhood's distribution of node attributes and architectural information. We use GraphSAGE to extract refined representations of nodes i.e., micro-video modalities and users. 
We sample the surrounding area around each node in the graph and use the message-passing procedure for information propagation and aggregation to learn the node embedding. 

The modality-specific nodes are initialized with the modality-specific representations as discussed in \autoref{ssec:fm}. The user nodes are initialized randomly. 
To acquire knowledge of the interplay between various categories of nodes, we use a message-passing paradigm. The message transfer technique is critical because it collects and merges the prominent aspects of neighboring nodes into the node embedding. Updating modality-specific embeddings through GraphSAGE enhances their representation learning by leveraging the graph structure and relationships among nodes. The initial modality-specific embeddings might capture low-level features, but they might lack contextual information. By updating these embeddings within the graph using GraphSAGE, we can incorporate contextual information and capture how different micro-videos are related to each other based on the similarity of their modality-specific features. This approach can lead to more informative, contextually aware embeddings, ultimately improving the performance of the task of hashtag recommendation for micro-videos. By using cosine similarity as edge weights, we capture semantic relationships between the same modalities of different micro-videos. This means visually similar or textually related micro-videos will have stronger connections in the graph. GraphSAGE can effectively propagate these relationships to refine and enhance the modality embeddings, enabling the model to capture the underlying semantics better. In contrast, the traditional deep learning methods focus solely on individual modalities and may not leverage such semantic relationships. Even though nodes of different modality types are not directly connected, GraphSAGE's ability to aggregate information from neighboring nodes can be advantageous. Within each modality type, we have connections between micro-videos, which means that similar micro-videos in terms of a specific modality can influence each other's embeddings. This allows the model to capture modality-specific nuances. While nodes of different modality types are not directly connected, the information can still propagate across modalities through transitive relationships. For example, if micro-video A is similar to micro-video B in terms of visual features, and micro-video B is similar to micro-video C in terms of textual features, GraphSAGE can potentially capture the indirect influence of visual features from micro-video A on textual features in micro-video C. Therefore, applying GraphSAGE to update modality embeddings extracted through deep learning tools enhances the model's ability to capture cross-modality relationships, semantic information, and contextual nuances.

\begin{algorithm}[!ht]
\caption{Feature Refinement}
\begin{tabular}{ll}
    \textit{Input:}&$G(N,E)$: Graph\\
    & $x_n, \forall n \in N$: Input features\\
    & $K$: Depth\\
    &$W^j, \forall j \in \{1,\hdots, J$\}: Weight matrices\\
    &$\sigma$: Non-linearity\\
    &$MEAN_j, \forall j \in \{1,\hdots, J\}$: Aggregator function\\
    &$F:n\rightarrow 2^N$: Neighborhood function\\
    \textit{Output:}&$z_n, \forall n \in N$: Vector representations\\
    \end{tabular}
\begin{algorithmic}[1]
    \STATE$h_n^0 \gets x_n, \forall n \in N$
     \FOR {$j=1$ to $J$}
        \FOR {$n \in N$}
            \STATE$h^j_n \gets \sigma(W^j . (\{h_n^{j-1}\} \cup MEAN_j( \{h_{n^{\prime}}^{j-1},\forall n^{\prime} \in F(n)\})))$
         \ENDFOR
    \STATE$h^j_n \gets h^j_n/ ||h^j_n||_2, \forall n \in N$
    \ENDFOR
    \STATE$z_n \gets h^J_n, \forall n \in N$
    \STATE$\textbf{return } z_n$
\end{algorithmic}
\label{alg:alg2} 
\end{algorithm}

The method for refining node embeddings is described in Algorithm~\autoref{alg:alg2}. The input consists of the whole graph, $G = (N, E)$, and feature vector representations for each node denoted by $x_n, \forall n \in N$. Here $x_n$ consists of visual features $mv^v_{initial}$, acoustic features $mv^a_{initial}$, textual features $mv^t_{initial}$, and user features $u_{initial}$. The for loop in Line 2 of Algorithm~\autoref{alg:alg2} runs for a total number of $J$ steps. Here $j$ denotes the current step and $h^j$ denotes a node's representation at $j^{th}$ step. The parameter $J$ regulates the method's neighborhood depth. If $J$ is 1, only the nearby nodes are recognized as similar. If $J$ is equal to 2, the nodes at distance two are also taken into account.
First, each node $n \in N$ accumulates representations of nodes in its near vicinity, $\{h^{j-1}_{n^{\prime}}, \forall n^{\prime} \in F(n)\}$, into a single neighborhood vector. Next, the node’s prior representation $h^{j-1}_n$ is combined with the consolidated neighborhood vector through concatenation. As shown in Line 4 of Algorithm~\autoref{alg:alg2}, we use the mean operator as our aggregator function to construct the node embedding.
where we compute the average of node embeddings in its neighborhood i.e., $\{h^{j-1}_{n^{\prime}}, \forall n^{\prime} \in F(n)\}$. The aforementioned vector is passed through a fully connected layer equipped with a non-linear activation function $\sigma$, which alters the representations intended for utilization in the subsequent phase of the algorithm (i.e., $h^j_n, \forall n \in N$). 
When each node is processed, we normalize the embeddings to have a unit norm as shown in Line 6 of Algorithm~\autoref{alg:alg2}. 
The final representations output at depth $J$ as $z_n = h^J_n, \forall n \in N$  are obtained as shown in~\autoref{eq:final}. 
\begin{equation}
z_n= \{mv^v_{final}, mv^a_{final}, mv^t_{final}, u_{final}\}
\label{eq:final}
\end{equation}
Here, ${mv^v_{final}, mv^a_{final}, mv^t_{final}, u_{final}}$ denotes the refined visual modality, acoustic modality, textual modality, and user feature vectors respectively.

\subsubsection{Micro-video Representation}
In the proposed framework, we jointly consider the modality-specific and user representations to investigate the impact that different modalities and users have on the overall micro-video representation. The content-based micro-video representation ($mv_{final}$) is obtained by concatenating modality-specific representations.
\begin{equation}
    mv_{final}=concat(mv^v_{final}, mv^a_{final}, mv^t_{final})
    \label{eq:eq12}
\end{equation}
We employed the concatenation operator since it helps to preserve the features in every modality. Subsequently, we concatenate the derived content-based micro-video embedding with user embedding to obtain the overall enriched micro-video representation as shown in~\autoref{eq:eq13}. 
\begin{equation}
mv_{overall}=concat(mv_{final},u_{final})
\label{eq:eq13}
\end{equation}
Here, $mv_{overall}$ is the derived micro-video representation. 
The hashtag recommendation module then uses this representation to anticipate hashtags for the given micro-video.
\subsection{Hashtag Recommendation}
The hashtag recommendation module takes the features extracted from the feature refinement module as input and yields a reasonable set of hashtag recommendations for a micro-video.  
Using the comprehensive feature vector ($mv_{overall}$) as input, we employ a dense layer of size $\lvert H \rvert$ followed by a softmax activation function to derive softmax scores for hashtags, as depicted in~\autoref{eq:eq14}.
\begin{equation}
    y_{pred} = softmax\Bigl(Dense\bigl(units = \lvert H \rvert \bigr) \bigl(mv_{overall}\bigr)\Bigr)
    \label{eq:eq14}
\end{equation}
Here, softmax probabilities of specified hashtags are represented by $y_{pred} \in R^{\lvert H \rvert}$. The final collection of anticipated hashtags is then obtained by using $argsort()$ that sorts hashtag according to softmax scores in descending order, as given in~\autoref{eq:eq15}.
\begin{equation}
    R= argsort(y_{pred})
\label{eq:eq15}
\end{equation}
Here, $R$ denotes the recommended hashtags.
 The training objective loss function is given in~\autoref{eq:eq16}.

\begin{equation}
    J=\frac{1}{\lvert Z\rvert}\sum_{( mv_i,G_i)\in Z}\sum_{g\in G_i}-log\Bigl( P\bigl( g|mv_i\bigr ) \Bigr)
     \label{eq:eq16}
\end{equation}
Here, $J$, is the loss function, $Z (Z \subset M)$ denotes the training set of micro-videos, $mv_i$ represents the current micro-video, $G_i$ denotes the corresponding ground-truth hashtag set, and
$P(g|mv_i)$ is the likelihood of selecting ground-truth hashtag ($g$) for the micro-video ($mv_i$).
\section{Experimental Evaluations}
\label{sec:expandresults}
To demonstrate the efficiency of our methodology, we first provide a description of the experimental conditions in this section, followed by the experimental findings.
\subsection{Experimental Setup}
In this section, we showcase various datasets utilized for conducting experiments. Afterward, we delve into distinct approaches employed for comparison, followed by evaluation metrics.
\subsubsection{Datasets}
We assess our devised framework on three real-world micro-video datasets namely, TMALL
~\citep{chen2016micro}, INSVIDEO~\citep{li2019long}, and YFCC~\citep{thomee2016yfcc100m}. 
\begin{table}[h]
\centering
\caption{Statistics of Different Datasets}\label{table:res1}
    \begin{tabular*}{0.9\linewidth}{@{\extracolsep{\fill}} lccccc }
    \toprule
\textbf{Datasets}& \textbf{{Micro-videos}} &\textbf{{Hashtags}}&\textbf{{Users}}& \textbf{$A_{h}$}&\textbf{$A_{mv}$}\\
\midrule
 TMALL~\citep{chen2016micro} & 13140 & 3354 & 839 & 44.24 & 15.66\\
 INSVIDEO~\citep{li2019long} & 30083 & 19930  & 2847 & 195.69 & 10.56  \\
 YFCC~\citep{thomee2016yfcc100m} & 16611 & 16354 & 1455 & 138.80 & 11.41 \\
\bottomrule
\end{tabular*}
\end{table}
We customized each dataset to match our needs for the task of hashtag recommendation for micro-videos. First, we conducted lemmatization on hashtags and later removed the low-frequency hashtags, i.e., hashtags appearing less than 50 times. Next, we removed those micro-videos that lacked any modality or hashtags. 
Further, we retain users who have posted at least four micro-video postings.~\autoref{table:res1} contains statistical information for all datasets after pre-processing. In~\autoref{table:res1}, $A_{mv}$ denotes the average number of micro-videos per user, and $A_h$ denotes the average number of hashtags per micro-video. 
TMALL, INSVIDEO, and YFCC datasets were collected from Vine\footnote{https://vine.co/}, Instagram, and Flickr\footnote{\label{f5}https://www.flickr.com/} platforms, respectively. 
For all these datasets, we take the train to test data split as 80:20.
Below, we go over these datasets in further detail.
\begin{itemize}
\item{TMALL}: 
\citep{chen2016micro} created this dataset for micro-video popularity prediction. Initially, there were 1.6 million video postings in the crawled dataset, including 3,03,242 distinct micro-videos with a combined runtime of 499.8 hours. After carrying out pre-processing steps, the dataset used in our research contained 13,140 micro-video posts and 3,354 distinct hashtags. The minimum, average, and maximum hashtag count per post is 4, 44.64, and 1,424, respectively. The dataset includes 839 unique individuals, each posting an average of 15.66 micro-videos. For every micro-video, the complete user profile and associated metadata are also available.
\item{INSVIDEO}:
\citep{li2019long} created the INSVIDEO dataset to advocate hashtags for micro-videos. The authors crawled micro-videos from Instagram with associated descriptions and hashtags. The crawled dataset contained 3,34,826 micro-videos and 9,170 users. The dataset used by 
\citep{li2019long} contains 2,13,847 micro-videos, 15,751 hashtags, and 6,786 users. Following pre-processing, the dataset contains 30,083 micro-video postings from 2,847 users, with a mean of 10.56 posts per user. The dataset contains micro-video posts with a range of hashtag counts, including a minimum of 4, an average of 13.4, and a maximum of 1,494.
\item{YFCC}:
The Yahoo Flickr Creative Commons 100M, dubbed as YFCC100M~\citep{thomee2016yfcc100m} dataset is a comprehensive publicly accessible multimodal dataset tht contains nearly 99.2 million photos and 0.8 million micro-videos from Flickr. To perform the task of micro-video hashtag recommendation, we crawled micro-videos, user profiles, and annotated hashtags. Finally, the collected dataset contained 1,34,992 micro-videos, 8,126 users, and 23,054 hashtags. Following data cleaning methods, the dataset used in our studies included 16,611 micro-videos and 16,354 unique hashtags, 1,455 unique users, and an average of 11.41 micro-videos per user. The micro-videos in the resulting dataset has a minimum of 4 and an average of 138.8 linked hashtags.
\end{itemize}
\subsubsection{Compared Methods}
In this section, we outline the prevailing models that recommend hashtags for micro-videos.
\begin{itemize}
\item Memory Augmented Co-attention Model (MACON)~\citep{zhang2019hashtag}: MACON is a hashtag recommendation system for multimodal content. The underlying premise of this approach is that texts and images play a critical role in providing information for tagging on social media platforms. The authors devise an attention mechanism that is mutually co-directed by both texts and images. It also acquires knowledge about the user's tagging preferences to provide customized recommendations. We used the implementation provided by the authors.
 \item User-Video Co-Attention Network (UVCAN)~\citep{liu2019user}: UVCAN was originally developed for personalized micro-video recommendation. UVCAN lays more emphasis on the user's hidden preference to obtain micro-video and user representation. UVCAN uses stacked attention techniques to learn multimodal information from both the user and micro-video. We have adapted for micro-video hashtag recommendation. 
\item Attention-based Multimodal Neural Network (AMNN)~\citep{yang2020amnn}: While attempting to generate hashtags in a sequence, 
\citep{yang2020amnn} employed an encoder-decoder architecture along with the softmax technique. The encoder uses CNN and Bi-LSTM to segregate the feature retrieval procedure for multimodal microblogs. The authors independently pay attention to each constituent modality and comprehend the fundamental components of images and captions. The concatenated visual and textual features are passed to GRU as input which then generates hashtags sequentially based on the probability scores.

\item Dual Graph Neural Network (DualGNN)~\citep{wang2021dualgnn}:
The two main modules that constitute DualGNN are single-modal and multimodal representation learning. The single-modal representation learning module uses the user-micro-video graph in each modality to identify unimodal user proclivities. In contrast, the multimodal representation learning module shows how the user weighs various modalities and infers the multimodal user preference. The ranking of the pertinent micro-videos for users is then done using a prediction mechanism. Initially designed for micro-video recommendation, this system is modified to recommend hashtags for micro-videos.

\item Learning the User’s Deeper Preferences (LUDP)~\citep{lei2023learning}:
A user-item interaction graph, an item-item modal similarity network, and a user preference graph for each modality are the three components that make up LUDP. Through the user-item interactions matrix, the authors construct a bipartite graph of users and items. 
The authors leverage modal information to propagate and aggregate item ID embeddings on the similarity network in order to generate modal similarity graphs and collect structural information about items. The multi-modal attributes are combined to represent the user's choice for the modal in the user preference graph, which is built based on the user's prior engagement with the item. These newly discovered user and item representations are combined with representations found through collaborative signals on the bipartite network to provide multimodal recommendations. We adapt LUDP to carry out hashtag recommendations for micro-videos.
\end{itemize}
\subsubsection{Evaluation Metrics}
To gauge the capability of our devised hashtag recommendation system, we use assessment criteria from the literature on multi-label classification.
The standard evaluation metrics for analysing how well hashtag recommendation systems perform are hit rate, precision, recall, and F1-score.  
These metrics are computed by comparing predicted hashtags and ground-truth hashtags for each micro-video post. 
Note that larger values indicate better performance.


\subsection{Experimental Results}
In this section, we outline the performance of our devised framework. We initially evaluate the performance of our proposed model against state-of-the-art methods on multiple datasets to determine its efficacy. Next, we analyze the performance gain, and performance of cold-start users, determine the sensitivity of various parameters, visualize the recommendations, and analyze the computational time. Note that $K$ denotes the number of suggested hashtags and that the findings in this section are expressed at $K=5$.
\subsubsection{Effectiveness Comparisons}
We undergo rigorous experiments on several datasets to highlight that our suggested model is superior to state-of-the-art methods.
\begin{itemize}
\item{Performance on TMALL Dataset}:
We assess the effectiveness of the proposed approach MISHON in comparison to its existing competitors.~\autoref{table:res2} highlights the experimental findings of our suggested technique against baselines on the TMALL dataset.
\begin{table}[!ht]
 \centering
        \caption{Effectiveness Comparison Results on TMALL Dataset}
        \label{table:res2}
    \begin{tabular*}{0.9\linewidth}{@{\extracolsep{\fill}} lcccc }
    \toprule
        \textbf{Methods} & \textbf{Hit rate} & \textbf{Precision} & \textbf{Recall} & \textbf{F1-score} \\         
    \midrule
 AMNN~\citep{yang2020amnn} & {0.374} & {0.127} & {0.268} & {0.172}\\
MACON~\citep{zhang2019hashtag} & 0.458 & 0.156 & 0.291 & 0.202    \\
UVCAN~\citep{liu2019user} & 0.586 & 0.165 & 0.355 & 0.225 \\
LUDP~\citep{lei2023learning} & {0.644}& {0.212}  & {0.432} & {0.284} \\
DUALGNN~\citep{wang2021dualgnn} & 0.707 & 0.257 & 0.505 &0.340\\
\textbf{MISHON}& {\textbf{0.753}} & {\textbf{0.283}} & {\textbf{0.563}}  & {\textbf{0.376}} \\
    \bottomrule   
    \end{tabular*}   
\end{table}
We can observe from~\autoref{table:res2} that MISHON outperforms the compared methods on the TMALL dataset. The relative improvement of our model in terms of hit rate, precision, recall, and F1-score is 37.9\%, 15.6\%, 29.5\%, 20.4\% over AMNN, and 29.5\%, 12.7\%, 27.2\%, 17.4\% over MACON. The performance improvement over AMNN is due to taking user correlations and users’ interactions with constituent modalities of posted micro-videos whereas AMNN solely considers the content information embedded in the post's multiple modalities. The reason behind the performance gain over MACON is that MISHON employs GraphSAGE to learn enriched embeddings of micro-video modalities and users whereas MACON relies on encoder-decoder architecture
coupled with a parallel co-attention mechanism. 
The relative improvement of MISHON is 16.7\%, 11.8\%, 20.8\%, 15.1\% over UVCAN, 10.9\%, 7.1\%, 13.1\%, and 9.2\% over LUDP, and 4.6\%, 2.6\%, 5.8\%, 3.6\% over DualGNN in terms of hit rate, precision, recall, and F1-score respectively. UVCAN does not take the acoustic modality into consideration. MISHON performs better than UVCAN due to the incorporation of three modalities constituting the micro-video. Although we mine collaborative information
of the user and modality-specific embeddings similar to DualGNN, MISHON achieves better performance. This is due to the inclusion of modality feature similarity, user correlations, and user interactions with constituent modalities of posted micro-videos. Unlike LUDP which creates three separate subgraphs, we create one graph containing four types of nodes and seven types of edges to enrich modality-specific representations based on semantic similarity, the user representations based on similar tagging behavior, user-to-modality interactions, and derive the embedding of the micro-video.

The performance comparison of hashtag recommendation models in terms of hit rate, precision, recall, and F1-score for a variable number of hashtags on the TMALL dataset is shown in~\autoref{fig:figres1}. Hit rate, precision, recall, and F1-score are plotted on the y-axis against the number of hashtags recommended on the x-axis. The count of suggested hashtags is between 1 and 9. As the number of suggested hashtags increases, hit rate and recall rise although precision drops. Our proposed model beats state-of-the-art models despite having a variable amount of suggested hashtags since its curves are consistently the highest across all performance criteria. The improvements in each of the four evaluation measures over extant methods demonstrate the capability and competitive advantage of our suggested model.
\begin{figure}[!ht]
\subfloat[Hit rate]{%
  \includegraphics[width=0.5\textwidth]{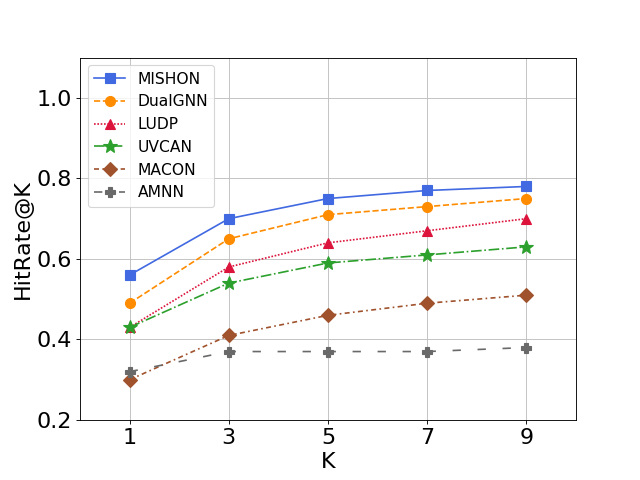}\label{fig:f1}%
}
\subfloat[Precision]{%
  \includegraphics[width=0.5\textwidth]{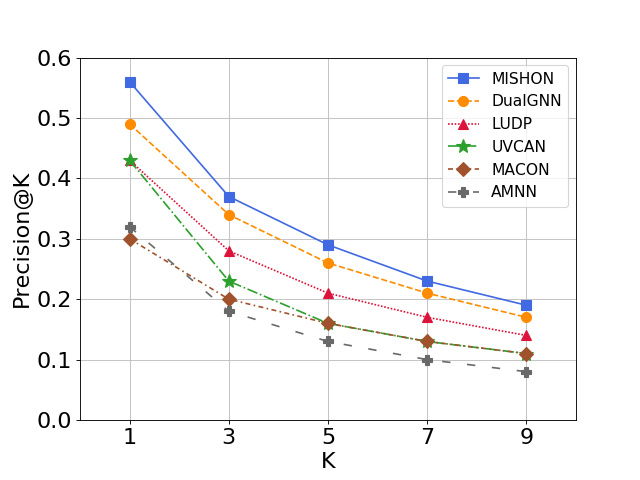}\label{fig:f2}%
}

\subfloat[Recall]{%
  \includegraphics[width=0.5\textwidth]{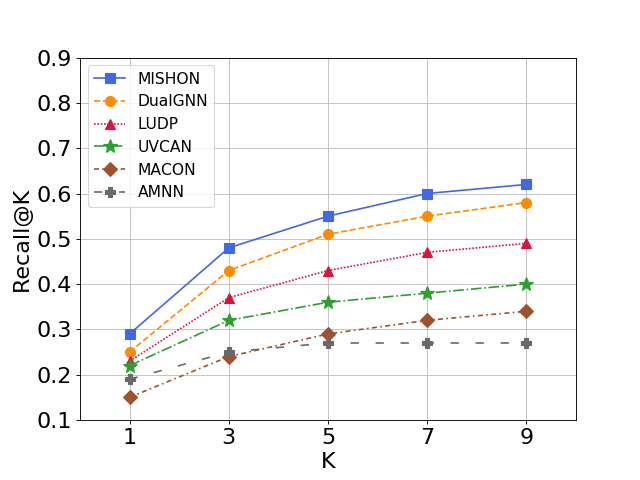}\label{fig:f3}%
}
\subfloat[F1-score]{%
  \includegraphics[width=0.5\textwidth]{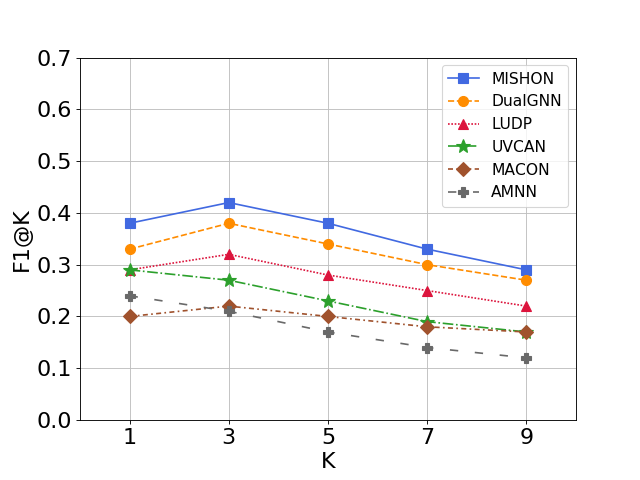}\label{fig:f4}%
}
\caption{Effectiveness Comparison Curves on TMALL Dataset}
\label{fig:figres1}
\end{figure}

\item{Performance on INSVIDEO Dataset}:
To investigate
the generalizability of MISHON in recommending hashtags
for micro-videos on different platforms, we experimented using the INSVIDEO dataset.  
\begin{table}[!b]
 \centering
 \caption{Effectiveness Comparison Results on INSVIDEO Dataset}
 \label{table:res3}
\begin{tabular*}{0.9\linewidth}{@{\extracolsep{\fill}} lcccc }
    \toprule
        \textbf{Methods} & \textbf{Hit rate} & \textbf{Precision} & \textbf{Recall} & \textbf{F1-score} \\         
    \midrule
 MACON~\citep{zhang2019hashtag} & 0.569 & 0.348 & 0.099 & 0.154    \\
UVCAN~\citep{liu2019user} & 0.747 & 0.407 & 0.132 & 0.200     \\
 AMNN~\citep{yang2020amnn} & {0.536} & {0.420} & {0.143} & {0.213}\\
 LUDP~\citep{lei2023learning} & {0.925}& {0.710}  & {0.253} & {0.373}  \\
DUALGNN~\citep{wang2021dualgnn} & 0.920 & 0.726 & 0.260 & 0.382\\
\textbf{MISHON}& {\textbf{0.941}} & {\textbf{0.764}} & {\textbf{0.280}}  & {\textbf{0.410}} \\
    \bottomrule   
    \end{tabular*}   
\end{table}
~\autoref{table:res3} shows the performance of our model against several other hashtag recommendation systems on INSVIDEO. Our model consistently outperforms AMNN by 40.5\%, 34.4\%, 13.7\%, and 19.7\%; MACON by 37.2\%, 41.6\%, 18.1\%, and 25.6\%; UVCAN by 19.4\%, 35.7\%, 14.8\%, and 21.0\%; LUDP by 1.6\%, 5.4\%, 2.7\%, 3.7\%;  and DualGNN by 2.1\%, 3.8\%, 2.0\%, and 2.8\% in terms of hit rate, precision, recall and F1-score respectively. We tend to see the same performance regime on the INSVIDEO dataset as observed in the case of the TMALL dataset.

\item{Performance on YFCC Dataset}:
We compare MISHON with other methods on the YFCC dataset to illustrate
its efficacy in micro-video hashtag recommendation.
\begin{table}[!ht]
 \centering
 \caption{Effectiveness Comparison Results on YFCC Dataset}
 \label{table:res4}
 \begin{tabular*}{0.9\linewidth}{@{\extracolsep{\fill}} lcccc }
    \toprule
        \textbf{Methods} & \textbf{Hit rate} & \textbf{Precision} & \textbf{Recall} & \textbf{F1-score} \\         
    \midrule
 MACON~\citep{zhang2019hashtag} & 0.465 & 0.218 & 0.164 & 0.187\\
 AMNN~\citep{yang2020amnn} & {0.527} & {0.330} & {0.284} & {0.305}\\
 UVCAN~\citep{liu2019user} & 0.543 & 0.188 & 0.176 & 0.182\\
 LUDP~\citep{lei2023learning} & {0.464}& {0.213}  & {0.171} & {0.190} \\
DUALGNN~\citep{wang2021dualgnn} & 
0.745 & 0.401 & 0.354 &  0.376\\
\textbf{MISHON}& \textbf{0.801} & {\textbf{0.471}} & {\textbf{0.413}}  & {\textbf{0.441}} \\
\bottomrule
\end{tabular*}
\end{table}

The performance of the suggested model is superior to that of state-of-the-art methods, as shown in~\autoref{table:res4}. In terms of hit rate, precision, recall, and F1-score, MISHON exhibits relative improvements of 27.4\%, 14.1\%, 12.9\%, and 13.6\% over AMNN; 33.6\%, 25.3\%, 24.9\%, and 25.4\% over MACON; 
25.8\%, 28.3\%, 23.7\%, and 25.9\% over UVCAN;
33.7\%, 25.8\%, 24.2\%, and 25.1\% over LUDP; and
5.6\%, 7.0\%, 5.9\%, and 6.5\% over DualGNN.
Our model generally maintains relative performance improvements across three datasets when compared to other approaches. The results demonstrate the superiority and effectiveness of our proposed method for recommending high-quality hashtags regardless of the platform taken into consideration.
\end{itemize}
\subsubsection{Performance Gain Analysis}
In this section, we examine how MISHON's performance improved as a result of the addition of the attention mechanism and the feature refinement module followed by various edge combinations. All experiments conducted in this section have been executed utilizing the TMALL dataset procured from the Vine platform.

\textit{Attention Mechanism:}
\begin{table}[!ht]
 \centering
 \caption{ Performance Comparison of Attention Mechanism}
 \label{table:res5}
    \begin{tabular*}{0.9\linewidth}{@{\extracolsep{\fill}} lcccc }
    \toprule
        \textbf{Methods} & \textbf{Hit rate} & \textbf{Precision} & \textbf{Recall} & \textbf{F1-score} \\         
    \midrule
Without Attention & {0.721}& {0.270}  & {0.527} & {0.357}  \\
\textbf{With Attention} & {\textbf{0.753}} & {\textbf{0.283}} & {\textbf{0.563}}  & {\textbf{0.376}} \\
 \bottomrule
 \end{tabular*}
\end{table}
We give experimental analyses in this section to illustrate the significance of the attention mechanism. In the absence of the attention mechanism from MISHON, we compute the average of the extracted modality-specific features and assign them as initial node embeddings.~\autoref{table:res5} shows the performance comparison yielded by the inclusion and exclusion of the attention mechanism. The performance improvement obtained by the inclusion of the attention mechanism is 3.2\%, 1.3\%, 3.6\%, and 1.9\% in hit rate, precision, recall, and F1-score respectively. The adopted attention mechanism determines the significance of each unit constituting the respective modalities and removes the unnecessary features. Typically, hashtags are used to draw attention to important details in micro-videos. Hashtags focus on key details more so than other information in the micro-video. An attention mechanism is used to determine the critical units in each modality in an adaptive manner that are most relevant to recommend hashtags. 
Since varied modalities have different representations, it is immensely important to assign differential weights to the information contained in the constituent modalities. This promising finding verifies the significance of using attention mechanisms in learning the important information contained in every modality in order to obtain the overall micro-video representation. 

\textit{Significance of Feature Refinement:}
We conducted experiments to highlight the significance of the feature refinement module. 
\begin{table}[!h]
\centering
\caption{Performance Comparison of of Feature Refinement Module}\label{table:res6}
  
       \begin{tabular*}{0.9\linewidth}{@{\extracolsep{\fill}} lcccc }
    \toprule
        \textbf{Methods} & \textbf{Hit rate} & \textbf{Precision} & \textbf{Recall} & \textbf{F1-score} \\         
    \midrule
Without FRM & {0.464}& {0.156}  & {0.303} & {0.207}  \\
\textbf{With FRM} & {\textbf{0.753}} & {\textbf{0.283}} & {\textbf{0.563}}  & {\textbf{0.376}} \\
    \bottomrule   
    \end{tabular*}   
\end{table}

~\autoref{table:res6} shows the performance comparison yielded by the inclusion and exclusion of the Feature Refinement Module (FRM). The performance improvement obtained when using node representations derived from FRM as opposed to without FRM is 28.9\%, 12.7\%, 26.0\%, and 16.9\% in hit rate, precision, recall, and F1-score respectively. We employ a graph neural network to leverage the graphical structure between users and micro-video modalities and produce informative node embeddings. Graph convolution operations capture both local structure information and the distribution of neighboring features.
The obtained modality-specific and user representations enrich the overall micro-video representation. Therefore, node embeddings in this study are derived from FRM using GraphSAGE.

\textit{Types of Edge Modeling:} We discuss the performance of our model with various edge modeling procedures below.
\begin{itemize}
    \item Homogeneous Edge Modeling: We construct intramodality edges i.e., edges between the same modality of different micro-videos and user-user edges with the criteria of modality similarity and user similarity respectively. 
    \item Heterogeneous Edge Modeling:
    We construct edges to connect nodes representing users and constituent modalities of their posted micro-videos to capture interrelationships, and high-order connectivity, in order to generate high-quality representations.

    \item Homogeneous + Heterogeneous Edge Modeling:
    We introduce edge generation procedures that simulate modality-to-modality, user-to-user, and user-to-modality interactions to focus on the important cues and suggest pertinent hashtags. 
\end{itemize}

\begin{table}[!ht]
\centering
\caption{Performance Comparison of Edge Modeling Techniques}\label{table:resem}
    \begin{tabular*}{0.9\linewidth}{@{\extracolsep{\fill}} lcccc }
    \toprule
        \textbf{Technique} & \textbf{Hit rate} & \textbf{Precision} & \textbf{Recall} & \textbf{F1-score} \\         
    \midrule
Homogeneous & {0.657}& {0.247}  & {0.457} & {0.320}  \\
Heterogeneous & {0.714}& {0.272}  & {0.513} & {0.353}  \\
\textbf{Both} & {\textbf{0.753}} & {\textbf{0.283}} & {\textbf{0.563}}  & {\textbf{0.376}} \\
\bottomrule
\end{tabular*}
\end{table}
~\autoref{table:resem} depicts the performance comparison with different edge modeling procedures. The performance improvement with both types of modeling is 9.6\%, 3.6\%, 10.6\%, 5.6\% over homogeneous modeling, and 3.9\%, 1.1\%, 5.0\%, 2.3\% over heterogeneous modeling in hit rate, precision, recall and F1-score respectively. 
We focus on important cues in micro-video modalities, hashtags, and users to model their interrelatedness. Our constructed graph encodes multimodal information of micro-videos and users into node features and captures their mutual influence to enrich their representations. The proposed model employing two distinct types of edges can fully exploit informative interactions among different nodes.
\subsubsection{Performance Analysis on Cold-Start Users}
In this section, we address the issue of hashtag recommendations for cold-start users.


\begin{table}[!ht]
\centering
\caption{Performance Comparison on Cold-Start Users}\label{table:res7}
    \begin{tabular*}{0.9\linewidth}{@{\extracolsep{\fill}} lcccc }
    \toprule
        \textbf{Technique} & \textbf{Hit rate} & \textbf{Precision} & \textbf{Recall} & \textbf{F1-score} \\         
    \midrule
MISHON (C) & {0.460}& {0.161}  & {0.315} & {0.213}  \\
\textbf{MISHON (SC)} & {\textbf{0.741}} & {\textbf{0.280}} & {\textbf{0.550}}  & {\textbf{0.371}} \\
\bottomrule
\end{tabular*}
\end{table}
~\autoref{table:res7} illustrates the capability of variants of MISHON in recommending hashtags for micro-videos posted on the Vine platform by cold-start users i.e., the TMALL dataset. Here, MISHON (C) utilizes only the micro-video content and MISHON (SC) employs the social influence technique besides content features to recommend hashtags for micro-videos posted by cold-start users. The performance gain of MISHON (SC) over MISHON (C) is 28.1\%, 11.9\%, 23.5\%, and 15.8\% in hit rate, precision, recall, and F1-score respectively. We speculate the performance improvement is due to modeling the influence of popular users on cold-start users. Users tend to follow hashtags used by the most popular users to gain social attention. We simulate the impact of social influence by applying GraphSAGE. MISHON (SC) employs the engagement rate of users for user-user edge construction. After information propagation and neighborhood aggregation, we can infer embeddings for cold-start users. 
\subsubsection{Parameter Sensitivity Studies}
We outline the sensitivity analysis of several model parameters: threshold value preprocessing, GraphSAGE aggregator function, thresholds for edge construction ($\gamma$  and $\theta$), popular users' selection ratio ($\alpha$), and the number of recommended hashtags.
\begin{itemize}
    \item Preprocessing Threshold: For preprocessing the low-frequency hashtags, we inspected them by varying the threshold in the range of 30 to 70. 
\begin{table}[!ht]
\centering
\caption{Performance of MISHON with Preprocessing Threshold}\label{table:respre}
    \begin{tabular*}{0.9\linewidth}{@{\extracolsep{\fill}} lcccc }
    \toprule
        \textbf{Threshold} & \textbf{Hit rate} & \textbf{Precision} & \textbf{Recall} & \textbf{F1-score} \\         
    \midrule
30                                    & 0.720             & 0.262              & 0.612           & 0.366             \\
40                                    & 0.712             & 0.251              & 0.613           & 0.356             \\
\textbf{50}                                    & \textbf{0.753}             & \textbf{0.283}              & \textbf{0.563}           & \textbf{0.376}             \\
60                                    & 0.704             & 0.247              & 0.622           & 0.353             \\
70                                    & 0.701             & 0.242              & 0.624           & 0.349 \\
\bottomrule
\end{tabular*}
\end{table}
    The results obtained on the TMALL dataset are tabulated in~\autoref{table:respre}. As can be seen from~\autoref{table:respre}, we obtain the maximum F1-score when the threshold is kept at 50. 
    Preprocessing by removing low-frequency hashtags is essential to reduce noise in the data. Low-frequency hashtags may introduce noise and make the model less robust. At a threshold of 50, we are retaining hashtags that occur at least 50 times. This ensures that we focus on relatively common and potentially informative hashtags while eliminating extremely rare or noisy ones. In real-world applications, it is crucial to find a balance that allows the model to be practical and efficient. A threshold of 50 is a practical choice because it ensures that the model is not overwhelmed by low-frequency noisy hashtags, essential in real-world scenarios where reliable performance is desirable. Thus, we filter out rare hashtags while focusing on the more consistent and informative ones. 
    \item Aggregator Functions: For aggregating neighborhood information, we experimented with various functions such as mean, max, min, sum, and product. 
The improvement of mean over other aggregator functions lies in the range of : [0.1-1.5]\%; [0.0-0.5]\%; [0.6-2.1]\%; and [0.1-0.7]\% in hit rate, precision, recall, and F1-score respectively.
This is because the mean aggregator function gathers data from neighboring nodes in its vicinity more comprehensively through the use of previous hashtag usage trends and user-posted micro-videos. Adopting the mean function helps to gather data from neighboring nodes more comprehensively and reduces noise in information propagation.
\item Effect of Threshold for Homogeneous Edge Filtration ($\theta$ and $\gamma$): We adjust values of threshold mentioned in~\autoref{eq:eq7} and~\autoref{eq:eq6} to filter out modality pairs and user pairs with low similarity. 
When $\theta$ and $\gamma$ are less than 0.5, performance suffers due to noisier, irrelevant modality pair and user-pair edges. However, when values are greater than 0.5, performance suffers because correlated neighbors can be neglected, implying that the ideal threshold is about 0.5.
\item Effect of Popular User Selection Ratio $(\alpha)$:
We run experiments to select the optimal ratio of popular users ($\alpha$). 
To get their content discovered on the platform and expand their audience, we assume that users tend to follow more well-known users. To this end, we first find the most popular users. Then we try to determine hashtags used by the most popular users to be given to cold-start users based on the content similarity. Popular users can be considered highly influential people and their hashtags are also adopted by other users to expand their social network and gain attention, and content visibility. Since hashtags are abstract labels to indicate topics, using popular hashtags related to that topic helps the micro-video posts created by cold-start or new users to be included under those categories. This usually results in gaining new followers and better reachability. To determine the association between users with varied engagement rates and cold-start users, this supposition is taken into account. In accordance with this supposition, we run experiments to find the optimal popular-user selection ratio ranging from 0.1 to 0.7. 
\begin{table}[!ht]
\centering
\caption{Sensitivity Analysis of Popular User Selection Ratio ($\alpha$)}\label{table:resalpha}
    \begin{tabular*}{0.9\linewidth}{@{\extracolsep{\fill}} lcccc }
    \toprule
        \textbf{$\alpha$} & \textbf{Hit rate} & \textbf{Precision} & \textbf{Recall} & \textbf{F1-score} \\         
    \midrule
 \textbf{0.1} & \textbf{0.741} & \textbf{0.280} & \textbf{0.548} & \textbf{0.371} \\
 0.3 & {0.736}& {0.278}  & {0.532} & {0.365}  \\
0.5 & {0.734} & {0.276} & {0.530} & {0.363}\\
 0.7 & 0.732 & 0.275 & 0.527 & 0.361 \\
\bottomrule
\end{tabular*}
\end{table}
As can be seen from~\autoref{table:resalpha}, the variations in the performance metrics (hit rate, precision, recall, and F1 score) are minimal and could be within the margin of error. This consistency suggests that the choice of $\alpha$ does not significantly impact the performance of the hashtag recommendation system for cold-start users. However, the best F1-score was obtained when $\alpha$ was set to 0.1. As we increase $\alpha$, the representations of users tend to be general rather than specific and inclined toward popular users. A selection ratio of 0.1 for popular users might be more realistic for many online platforms. In practice, only a small percentage of users tend to be extremely popular. This aligns with real-world usage patterns on social media platforms. We aim to build a recommendation system that performs well across different platforms or domains; a conservative selection of popular users (0.1) may provide a better generalization to various contexts than higher $\alpha$ values. This aligns with realistic usage patterns and provides stability and consistency in results.
\item Number of Recommended Hashtags:
To examine the effectiveness of hashtag recommendation methods, we employ various values of $K$. The outcomes for $K$ = 1 to 9, representing the number of top-ranked recommended hashtags, are illustrated in~\autoref{fig:figres1}. We specifically set $K$ = 5 in our work.
\end{itemize}
Overall, the proposed technique is reliable throughout a significant range of these parameters. 
In summary, we set the preprocessing threshold to 50, aggregator function to mean, $\gamma$, and $\theta$ to 0.5, $\alpha$=0.1, and K=5 in order to achieve the results that we report in the paper.
\subsubsection{Qualitative Analysis}
\begin{figure}[!ht]
	\centering
        \includegraphics[width=\textwidth]{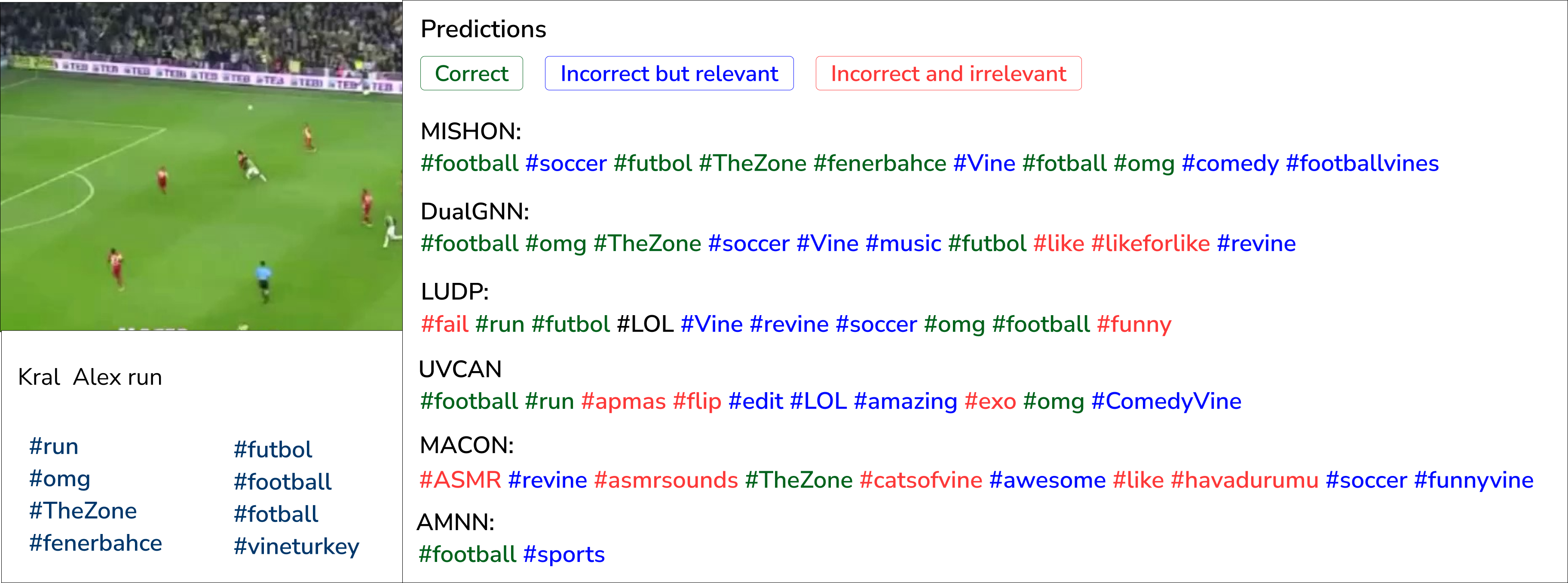}
	\caption{Example Post}
	\label{fig:post1}
\end{figure}

To visually depict the quality of hashtags suggested by several methods, we present a micro-video post sourced from the Vine platform in~\autoref{fig:post1}.
This example post has been chosen from test data, with correct, relevant, and incorrect hashtags shown in green, blue, and red respectively. The recommended hashtags matching ground-truth hashtags are called correct, relevant hashtags are consistent with the micro-video content but not specified in the set of ground-truth hashtags, and incorrect hashtags are neither correct nor relevant. As can be seen in~\autoref{fig:post1}, our model recommends the highest number of correct hashtags as opposed to four, four, three, one, and one hashtags recommended by DualGNN, LUDP, UVCAN, MACON, and AMNN respectively. MISHON recommends \#vine and \#footballvines which are logical recommendations since this post is related to football and is uploaded on Vine. Furthermore, our model recommends \#comedy. This hashtag is deemed relevant because it has been derived from the acoustic modality of the micro-video shown in the example post. This justifies the importance of mining information from constituent modalities of micro-videos. 
MISHON utilizes multimodal content information, user correlations, and interactions to recommend hashtags for micro-videos. The higher the quality of hashtags recommended, the more likely users are to assign hashtags to micro-videos, thus enriching the user experience. 
As a result of better hashtag recommendations by our proposed model, more people will enter hashtag channels they actually enjoy and spend more time viewing hashtag-specific micro-videos. 

\subsubsection{Time Analysis}
In this section, we analyze the execution time of different models. The time taken by different models per epoch tabulated in~\autoref{table:time} is reported in seconds. The experiments were conducted on a Linux Server featuring an Intel(R) Xeon(R) Silver 4215R CPU operating at 3.20 GHz, equipped with 256 GB RAM, and a 16-GB NVIDIA Tesla T4 GPU. 
\begin{table}[!ht]
\centering
\caption{Computational Time}\label{table:time}
    \begin{tabular*}{0.9\linewidth}{@{\extracolsep{\fill}} lcccc }
    \toprule
{\textbf{Model}}& \textbf{TMALL} & \textbf{Long tail} & \textbf{YFCC}\\
\midrule
DUALGNN~\citep{wang2021dualgnn} & {13}& {36} & {14} \\
UVCAN~\citep{liu2019user} & 44 &20 &18     \\
LUDP~\citep{lei2023learning} & 27 & 26 & 26\\
MACON~\citep{zhang2019hashtag} & 100 & 270 & 270\\
AMNN~\citep{yang2020amnn} & 469 & 1284 & 507\\
MISHON& 479 & 1198 & 669\\
\bottomrule
\end{tabular*}
\end{table}
As evident from~\autoref{table:time}, our proposed model, i.e., MISHON, takes the maximum time on three datasets in comparison to its counterparts in the context of hashtag recommendation. This disparity arises from the intricate architecture of our model which is composed of graph neural networks, Transfer Learning, and Deep Learning-based models. The incorporation of graph neural networks is instrumental in simulating interactions between modalities, users, and the interplay between users and modalities. This approach effectively captures the mutual influence exerted by constituent modalities within micro-videos and corresponding users. 

Although our model consumes more resources, its temporal performance aligns with that of other models. Despite the increased computational overhead, MISHON delivers superior quantitative and qualitative performance across all datasets. This also highlights the fact that our model can easily adapt to different platforms, each with its own user base and content characteristics. Furthermore, MISHON recommends a rich quality of relevant and personalized hashtags that encapsulate not only the multifaceted content information but also the user’s hashtagging preferences and broader community impact.

\section{Discussion}
\label{sec:discussion}
Our proposed method considers the shared interests of a user with like-minded users, the user's personal preferences, and micro-video content and allows for a more personalized user experience by suggesting relevant hashtags. By analyzing user behavior, i.e., user preferences and interactions, hashtags can be recommended based on the collective wisdom of similar users. Moreover, integrating content-based features further tailors recommendations to match individual preferences and align with attributes of micro-videos that users have shown interest in. Collectively, these approaches help in recommending high-quality and relevant hashtags for micro-videos. This highlights the benefits of hybridization in recommendation systems that can significantly enhance the system's accuracy, personalization, and scalability. Using graph neural networks to enrich the semantic depiction of user and modality-specific nodes can enhance their representations, as nodes with similar meanings tend to congregate in the neighboring semantic realm. The proposed social influence and content-based solutions can address the cold-start user problem. This solution allows the system to make recommendations for cold-start users by modeling their interaction with influential users and discovering popular hashtagging trends, even in the absence of the user’s historical tagging data. This assists in overcoming the initial data scarcity issue and yields meaningful hashtag recommendations.

The proposed hashtag recommendation system for micro-videos has several implications that can enhance the user experience and content discovery. By leveraging the synergies between user behavior and content characteristics, MISHON can deliver highly relevant and engaging hashtag recommendations to users, ultimately improving their overall micro-video viewing experience. With significant performance improvements, MISHON can facilitate community building by suggesting relevant hashtags that connect users with similar interests and stimulate participation in hashtag-based discussions on the micro-video platform. This benefits the platform by driving user activity and retention. More users are encouraged to explore videos in their areas of interest, increasing video interaction and resulting in a more engaging and interactive user community.

\section{Conclusion}
\label{sec:conclusions}
This study aims to recommend pertinent hashtags for micro-videos while also alleviating the cold-start user problem. We propose effective hybrid filtering for micro-video hashtag recommendation based on Deep Learning and GNN. The proposed framework comprises three components: feature mining; feature refinement; and hashtag recommendation. The feature mining module attentively derives features from modalities constituting micro-videos. In the feature refinement module, we construct a graph using the constituent modalities of micro-videos and corresponding users as nodes. The edges are built to simulate modality-to-modality, user-to-user, and user-to-modality interactions. The user representation is derived by inductively modeling the hashtag preferences of like-minded users. The constructed graph enables learning of high-quality node embeddings based on information propagation and neighborhood aggregation. We run comprehensive experiments on three real-world datasets comprising users’ posted micro-videos with accompanying hashtags.  We also alleviate the cold-start user problem by proposing a social influence and content-based technique to yield hashtags for micro-videos posted by them. Our proposed approach demonstrates superior performance compared to the existing methods.

Expanding upon the achievements of our current study, we contemplate the exploration of the subsequent avenues for prospective research. Our current model (MISHON) considers the static content of micro-videos when generating hashtag recommendations. Nevertheless, micro-videos exhibit topicality and temporal sensitivity, implying that the relevance of hashtags associated with a given micro-video can evolve over time. In subsequent research, we intend to harness the temporal dimension inherent in micro-videos to yield recommendations that are more temporally aligned and pertinent. To further enhance MISHON, we envisage the incorporation of a broader spectrum of user-centric contextual information. This encompasses diverse factors such as user profiles, demographic attributes, and geographical location data. 

\bibliography{08_bibliography}

\begin{thebibliography}{44}
\expandafter\ifx\csname natexlab\endcsname\relax\def\natexlab#1{#1}\fi
\providecommand{\url}[1]{\texttt{#1}}
\providecommand{\href}[2]{#2}
\providecommand{\path}[1]{#1}
\providecommand{\DOIprefix}{doi:}
\providecommand{\ArXivprefix}{arXiv:}
\providecommand{\URLprefix}{URL: }
\providecommand{\Pubmedprefix}{pmid:}
\providecommand{\doi}[1]{\href{http://dx.doi.org/#1}{\path{#1}}}
\providecommand{\Pubmed}[1]{\href{pmid:#1}{\path{#1}}}
\providecommand{\bibinfo}[2]{#2}
\ifx\xfnm\relax \def\xfnm[#1]{\unskip,\space#1}\fi
\bibitem[{Alsini et~al.(2020)Alsini, Datta \& Huynh}]{alsini2020utilizing}
\bibinfo{author}{Alsini, A.}, \bibinfo{author}{Datta, A.}, \& \bibinfo{author}{Huynh, D.~Q.} (\bibinfo{year}{2020}).
\newblock \bibinfo{title}{On utilizing communities detected from social networks in hashtag recommendation}.
\newblock {\it \bibinfo{journal}{IEEE Transactions on Computational Social Systems}\/},  {\it \bibinfo{volume}{7}\/}, \bibinfo{pages}{971--982}.
\bibitem[{Baevski et~al.(2020)Baevski, Zhou, Mohamed \& Auli}]{baevski2020wav2vec}
\bibinfo{author}{Baevski, A.}, \bibinfo{author}{Zhou, Y.}, \bibinfo{author}{Mohamed, A.}, \& \bibinfo{author}{Auli, M.} (\bibinfo{year}{2020}).
\newblock \bibinfo{title}{wav2vec 2.0: A framework for self-supervised learning of speech representations}.
\newblock {\it \bibinfo{journal}{Advances in Neural Information Processing Systems}\/},  {\it \bibinfo{volume}{33}\/}, \bibinfo{pages}{12449--12460}.
\bibitem[{Bansal et~al.(2022)Bansal, Gowda \& Kumar}]{bansal2022hybrid}
\bibinfo{author}{Bansal, S.}, \bibinfo{author}{Gowda, K.}, \& \bibinfo{author}{Kumar, N.} (\bibinfo{year}{2022}).
\newblock \bibinfo{title}{A hybrid deep neural network for multimodal personalized hashtag recommendation}.
\newblock {\it \bibinfo{journal}{IEEE transactions on computational social systems}\/}, .
\bibitem[{Bansal et~al.(2024)Bansal, Gowda \& Kumar}]{bansal2024multilingual}
\bibinfo{author}{Bansal, S.}, \bibinfo{author}{Gowda, K.}, \& \bibinfo{author}{Kumar, N.} (\bibinfo{year}{2024}).
\newblock \bibinfo{title}{Multilingual personalized hashtag recommendation for low resource indic languages using graph-based deep neural network}.
\newblock {\it \bibinfo{journal}{Expert Systems with Applications}\/},  {\it \bibinfo{volume}{236}\/}, \bibinfo{pages}{121188}.
\bibitem[{Basch et~al.(2021)Basch, Fera, Pellicane \& Basch}]{basch2021videos}
\bibinfo{author}{Basch, C.~H.}, \bibinfo{author}{Fera, J.}, \bibinfo{author}{Pellicane, A.}, \& \bibinfo{author}{Basch, C.~E.} (\bibinfo{year}{2021}).
\newblock \bibinfo{title}{Videos with the hashtag\# vaping on tiktok and implications for informed decision-making by adolescents: descriptive study}.
\newblock {\it \bibinfo{journal}{JMIR Pediatrics and Parenting}\/},  {\it \bibinfo{volume}{4}\/}, \bibinfo{pages}{e30681}.
\bibitem[{Bel{\'e}m et~al.(2019)Bel{\'e}m, Heringer, Almeida \& Gon{\c{c}}alves}]{belem2019exploiting}
\bibinfo{author}{Bel{\'e}m, F.~M.}, \bibinfo{author}{Heringer, A.~G.}, \bibinfo{author}{Almeida, J.~M.}, \& \bibinfo{author}{Gon{\c{c}}alves, M.~A.} (\bibinfo{year}{2019}).
\newblock \bibinfo{title}{Exploiting syntactic and neighbourhood attributes to address cold start in tag recommendation}.
\newblock {\it \bibinfo{journal}{Information Processing \& Management}\/},  {\it \bibinfo{volume}{56}\/}, \bibinfo{pages}{771--790}.
\bibitem[{Cantini et~al.(2021)Cantini, Marozzo, Bruno \& Trunfio}]{cantini2021learning}
\bibinfo{author}{Cantini, R.}, \bibinfo{author}{Marozzo, F.}, \bibinfo{author}{Bruno, G.}, \& \bibinfo{author}{Trunfio, P.} (\bibinfo{year}{2021}).
\newblock \bibinfo{title}{Learning sentence-to-hashtags semantic mapping for hashtag recommendation on microblogs}.
\newblock {\it \bibinfo{journal}{ACM Transactions on Knowledge Discovery from Data (TKDD)}\/},  {\it \bibinfo{volume}{16}\/}, \bibinfo{pages}{1--26}.
\bibitem[{Cao et~al.(2020)Cao, Miao, Rong, Qin \& Nie}]{cao2020hashtag}
\bibinfo{author}{Cao, D.}, \bibinfo{author}{Miao, L.}, \bibinfo{author}{Rong, H.}, \bibinfo{author}{Qin, Z.}, \& \bibinfo{author}{Nie, L.} (\bibinfo{year}{2020}).
\newblock \bibinfo{title}{Hashtag our stories: Hashtag recommendation for micro-videos via harnessing multiple modalities}.
\newblock {\it \bibinfo{journal}{Knowledge-Based Systems}\/},  {\it \bibinfo{volume}{203}\/}, \bibinfo{pages}{106114}.
\bibitem[{Chakrabarti et~al.(2023)Chakrabarti, Malvi, Bansal \& Kumar}]{chakrabarti2023hashtag}
\bibinfo{author}{Chakrabarti, P.}, \bibinfo{author}{Malvi, E.}, \bibinfo{author}{Bansal, S.}, \& \bibinfo{author}{Kumar, N.} (\bibinfo{year}{2023}).
\newblock \bibinfo{title}{Hashtag recommendation for enhancing the popularity of social media posts}.
\newblock {\it \bibinfo{journal}{Social Network Analysis and Mining}\/},  {\it \bibinfo{volume}{13}\/}, \bibinfo{pages}{21}.
\bibitem[{Chen et~al.(2016)Chen, Song, Nie, Wang, Zhang \& Chua}]{chen2016micro}
\bibinfo{author}{Chen, J.}, \bibinfo{author}{Song, X.}, \bibinfo{author}{Nie, L.}, \bibinfo{author}{Wang, X.}, \bibinfo{author}{Zhang, H.}, \& \bibinfo{author}{Chua, T.-S.} (\bibinfo{year}{2016}).
\newblock \bibinfo{title}{Micro tells macro: Predicting the popularity of micro-videos via a transductive model}.
\newblock In {\it \bibinfo{booktitle}{Proceedings of the 24th ACM international conference on Multimedia}\/} (pp. \bibinfo{pages}{898--907}).
\bibitem[{Chen et~al.(2021)Chen, Lai, Liu \& Chen}]{chen2021tagnet}
\bibinfo{author}{Chen, Y.-C.}, \bibinfo{author}{Lai, K.-T.}, \bibinfo{author}{Liu, D.}, \& \bibinfo{author}{Chen, M.-S.} (\bibinfo{year}{2021}).
\newblock \bibinfo{title}{Tagnet: triplet-attention graph networks for hashtag recommendation}.
\newblock {\it \bibinfo{journal}{IEEE Transactions on Circuits and Systems for Video Technology}\/},  {\it \bibinfo{volume}{32}\/}, \bibinfo{pages}{1148--1159}.
\bibitem[{Cho et~al.(2014)Cho, van Merri{\"e}nboer, Gu̇l{\c{c}}ehre, Bahdanau, Bougares, Schwenk \& Bengio}]{cho2014learning}
\bibinfo{author}{Cho, K.}, \bibinfo{author}{van Merri{\"e}nboer, B.}, \bibinfo{author}{Gu̇l{\c{c}}ehre, {\c{C}}.}, \bibinfo{author}{Bahdanau, D.}, \bibinfo{author}{Bougares, F.}, \bibinfo{author}{Schwenk, H.}, \& \bibinfo{author}{Bengio, Y.} (\bibinfo{year}{2014}).
\newblock \bibinfo{title}{Learning phrase representations using rnn encoder--decoder for statistical machine translation}.
\newblock In {\it \bibinfo{booktitle}{Proceedings of the 2014 Conference on Empirical Methods in Natural Language Processing (EMNLP)}\/} (pp. \bibinfo{pages}{1724--1734}).
\bibitem[{Djenouri et~al.(2022{\natexlab{a}})Djenouri, Belhadi, Srivastava \& Lin}]{djenouri2022deep}
\bibinfo{author}{Djenouri, Y.}, \bibinfo{author}{Belhadi, A.}, \bibinfo{author}{Srivastava, G.}, \& \bibinfo{author}{Lin, J. C.-W.} (\bibinfo{year}{2022}{\natexlab{a}}).
\newblock \bibinfo{title}{Deep learning based hashtag recommendation system for multimedia data}.
\newblock {\it \bibinfo{journal}{Information Sciences}\/},  {\it \bibinfo{volume}{609}\/}, \bibinfo{pages}{1506--1517}.
\bibitem[{Djenouri et~al.(2022{\natexlab{b}})Djenouri, Belhadi, Srivastava \& Lin}]{djenouri2022toward}
\bibinfo{author}{Djenouri, Y.}, \bibinfo{author}{Belhadi, A.}, \bibinfo{author}{Srivastava, G.}, \& \bibinfo{author}{Lin, J. C.-W.} (\bibinfo{year}{2022}{\natexlab{b}}).
\newblock \bibinfo{title}{Toward a cognitive-inspired hashtag recommendation for twitter data analysis}.
\newblock {\it \bibinfo{journal}{IEEE Transactions on Computational Social Systems}\/},  {\it \bibinfo{volume}{9}\/}, \bibinfo{pages}{1748--1757}.
\bibitem[{Dosovitskiy et~al.(2021)Dosovitskiy, Beyer, Kolesnikov, Weissenborn, Zhai, Unterthiner, Dehghani, Minderer, Heigold, Gelly, Uszkoreit \& Houlsby}]{dosovitskiy2021an}
\bibinfo{author}{Dosovitskiy, A.}, \bibinfo{author}{Beyer, L.}, \bibinfo{author}{Kolesnikov, A.}, \bibinfo{author}{Weissenborn, D.}, \bibinfo{author}{Zhai, X.}, \bibinfo{author}{Unterthiner, T.}, \bibinfo{author}{Dehghani, M.}, \bibinfo{author}{Minderer, M.}, \bibinfo{author}{Heigold, G.}, \bibinfo{author}{Gelly, S.}, \bibinfo{author}{Uszkoreit, J.}, \& \bibinfo{author}{Houlsby, N.} (\bibinfo{year}{2021}).
\newblock \bibinfo{title}{An image is worth 16x16 words: Transformers for image recognition at scale}.
\newblock In {\it \bibinfo{booktitle}{International Conference on Learning Representations}\/}.
\newblock \URLprefix \url{https://openreview.net/forum?id=YicbFdNTTy}.
\bibitem[{Duan et~al.(2023)Duan, Zhu, Liang, Zhu \& Liu}]{duan2023multi}
\bibinfo{author}{Duan, H.}, \bibinfo{author}{Zhu, Y.}, \bibinfo{author}{Liang, X.}, \bibinfo{author}{Zhu, Z.}, \& \bibinfo{author}{Liu, P.} (\bibinfo{year}{2023}).
\newblock \bibinfo{title}{Multi-feature fused collaborative attention network for sequential recommendation with semantic-enriched contrastive learning}.
\newblock {\it \bibinfo{journal}{Information Processing \& Management}\/},  {\it \bibinfo{volume}{60}\/}, \bibinfo{pages}{103416}.
\bibitem[{Hamilton et~al.(2017)Hamilton, Ying \& Leskovec}]{hamilton2017inductive}
\bibinfo{author}{Hamilton, W.}, \bibinfo{author}{Ying, Z.}, \& \bibinfo{author}{Leskovec, J.} (\bibinfo{year}{2017}).
\newblock \bibinfo{title}{Inductive representation learning on large graphs}.
\newblock {\it \bibinfo{journal}{Advances in neural information processing systems}\/},  {\it \bibinfo{volume}{30}\/}.
\bibitem[{He et~al.(2023)He, Zhang, Zhang \& Ren}]{he2023usbe}
\bibinfo{author}{He, H.}, \bibinfo{author}{Zhang, R.}, \bibinfo{author}{Zhang, Y.}, \& \bibinfo{author}{Ren, J.} (\bibinfo{year}{2023}).
\newblock \bibinfo{title}{Usbe: User-similarity based estimator for multimedia cold-start recommendation}.
\newblock {\it \bibinfo{journal}{Multimedia Tools and Applications}\/},  (pp. \bibinfo{pages}{1--16}).
\bibitem[{Jafari et~al.(2023)Jafari, Luo \& Jafari}]{jafari2023unsupervised}
\bibinfo{author}{Jafari, B.~M.}, \bibinfo{author}{Luo, X.}, \& \bibinfo{author}{Jafari, A.} (\bibinfo{year}{2023}).
\newblock \bibinfo{title}{Unsupervised keyword extraction for hashtag recommendation in social media}.
\newblock In {\it \bibinfo{booktitle}{The International FLAIRS Conference Proceedings}\/}.
\newblock volume~\bibinfo{volume}{36}.
\bibitem[{Jafari~Sadr et~al.(2023)Jafari~Sadr, Mirtaheri, Greco, Borna et~al.}]{jafari2023popular}
\bibinfo{author}{Jafari~Sadr, M.}, \bibinfo{author}{Mirtaheri, S.~L.}, \bibinfo{author}{Greco, S.}, \bibinfo{author}{Borna, K.} et~al. (\bibinfo{year}{2023}).
\newblock \bibinfo{title}{Popular tag recommendation by neural network in social media}.
\newblock {\it \bibinfo{journal}{Computational Intelligence and Neuroscience}\/},  {\it \bibinfo{volume}{2023}\/}.
\bibitem[{Kenton \& Toutanova(2019)}]{kenton2019bert}
\bibinfo{author}{Kenton, J. D. M.-W.~C.}, \& \bibinfo{author}{Toutanova, L.~K.} (\bibinfo{year}{2019}).
\newblock \bibinfo{title}{Bert: Pre-training of deep bidirectional transformers for language understanding}.
\newblock In {\it \bibinfo{booktitle}{Proceedings of NAACL-HLT}\/} (pp. \bibinfo{pages}{4171--4186}).
\bibitem[{Kumar et~al.(2021)Kumar, Baskaran, Konjengbam \& Singh}]{kumar2021hashtag}
\bibinfo{author}{Kumar, N.}, \bibinfo{author}{Baskaran, E.}, \bibinfo{author}{Konjengbam, A.}, \& \bibinfo{author}{Singh, M.} (\bibinfo{year}{2021}).
\newblock \bibinfo{title}{Hashtag recommendation for short social media texts using word-embeddings and external knowledge}.
\newblock {\it \bibinfo{journal}{Knowledge and Information Systems}\/},  {\it \bibinfo{volume}{63}\/}, \bibinfo{pages}{175--198}.
\bibitem[{Lakshmi et~al.(2023)Lakshmi, Gerard, Santhanavijayan \& Radha}]{lakshmi2023hybrid}
\bibinfo{author}{Lakshmi, R.~V.}, \bibinfo{author}{Gerard, D.}, \bibinfo{author}{Santhanavijayan, A.}, \& \bibinfo{author}{Radha, S.} (\bibinfo{year}{2023}).
\newblock \bibinfo{title}{A hybrid classifier-based ontology driven image tag recommendation framework for social image tagging}.
\newblock {\it \bibinfo{journal}{Procedia Computer Science}\/},  {\it \bibinfo{volume}{218}\/}, \bibinfo{pages}{67--73}.
\bibitem[{Lei et~al.(2023)Lei, Cao, Yang, Ding \& Zhang}]{lei2023learning}
\bibinfo{author}{Lei, F.}, \bibinfo{author}{Cao, Z.}, \bibinfo{author}{Yang, Y.}, \bibinfo{author}{Ding, Y.}, \& \bibinfo{author}{Zhang, C.} (\bibinfo{year}{2023}).
\newblock \bibinfo{title}{Learning the user’s deeper preferences for multi-modal recommendation systems}.
\newblock {\it \bibinfo{journal}{ACM Transactions on Multimedia Computing, Communications and Applications}\/},  {\it \bibinfo{volume}{19}\/}, \bibinfo{pages}{1--18}.
\bibitem[{Li et~al.(2019)Li, Gan, Liu, Cheng, Yin \& Nie}]{li2019long}
\bibinfo{author}{Li, M.}, \bibinfo{author}{Gan, T.}, \bibinfo{author}{Liu, M.}, \bibinfo{author}{Cheng, Z.}, \bibinfo{author}{Yin, J.}, \& \bibinfo{author}{Nie, L.} (\bibinfo{year}{2019}).
\newblock \bibinfo{title}{Long-tail hashtag recommendation for micro-videos with graph convolutional network}.
\newblock In {\it \bibinfo{booktitle}{Proceedings of the 28th ACM International Conference on Information and Knowledge Management}\/} (pp. \bibinfo{pages}{509--518}).
\bibitem[{Liu et~al.(2019)Liu, Chen, Liu \& Hu}]{liu2019user}
\bibinfo{author}{Liu, S.}, \bibinfo{author}{Chen, Z.}, \bibinfo{author}{Liu, H.}, \& \bibinfo{author}{Hu, X.} (\bibinfo{year}{2019}).
\newblock \bibinfo{title}{User-video co-attention network for personalized micro-video recommendation}.
\newblock In {\it \bibinfo{booktitle}{The World Wide Web Conference}\/} (pp. \bibinfo{pages}{3020--3026}).
\bibitem[{Liu et~al.(2020)Liu, Xie, Zou \& Chen}]{liu2020user}
\bibinfo{author}{Liu, S.}, \bibinfo{author}{Xie, J.}, \bibinfo{author}{Zou, C.}, \& \bibinfo{author}{Chen, Z.} (\bibinfo{year}{2020}).
\newblock \bibinfo{title}{User conditional hashtag recommendation for micro-videos}.
\newblock In {\it \bibinfo{booktitle}{2020 IEEE International Conference on Multimedia and Expo (ICME)}\/} (pp. \bibinfo{pages}{1--6}).
\newblock \bibinfo{organization}{IEEE}.
\bibitem[{Mehta et~al.(2021)Mehta, Sarkhel, Chen, Mitra, Swaminathan, Rossi, Aminian, Guo \& Garg}]{mehta2021open}
\bibinfo{author}{Mehta, S.}, \bibinfo{author}{Sarkhel, S.}, \bibinfo{author}{Chen, X.}, \bibinfo{author}{Mitra, S.}, \bibinfo{author}{Swaminathan, V.}, \bibinfo{author}{Rossi, R.}, \bibinfo{author}{Aminian, A.}, \bibinfo{author}{Guo, H.}, \& \bibinfo{author}{Garg, K.} (\bibinfo{year}{2021}).
\newblock \bibinfo{title}{Open-domain trending hashtag recommendation for videos}.
\newblock In {\it \bibinfo{booktitle}{2021 IEEE International Symposium on Multimedia (ISM)}\/} (pp. \bibinfo{pages}{174--181}).
\newblock \bibinfo{organization}{IEEE}.
\bibitem[{Panayotov et~al.(2015)Panayotov, Chen, Povey \& Khudanpur}]{panayotov2015librispeech}
\bibinfo{author}{Panayotov, V.}, \bibinfo{author}{Chen, G.}, \bibinfo{author}{Povey, D.}, \& \bibinfo{author}{Khudanpur, S.} (\bibinfo{year}{2015}).
\newblock \bibinfo{title}{Librispeech: an asr corpus based on public domain audio books}.
\newblock In {\it \bibinfo{booktitle}{2015 IEEE international conference on acoustics, speech and signal processing (ICASSP)}\/} (pp. \bibinfo{pages}{5206--5210}).
\newblock \bibinfo{organization}{IEEE}.
\bibitem[{ur~Rehman et~al.(2023)ur~Rehman, Ali, Jan, Ali, Xu \& Shao}]{ur2023caml}
\bibinfo{author}{ur~Rehman, I.}, \bibinfo{author}{Ali, W.}, \bibinfo{author}{Jan, Z.}, \bibinfo{author}{Ali, Z.}, \bibinfo{author}{Xu, H.}, \& \bibinfo{author}{Shao, J.} (\bibinfo{year}{2023}).
\newblock \bibinfo{title}{Caml: Contextual augmented meta-learning for cold-start recommendation}.
\newblock {\it \bibinfo{journal}{Neurocomputing}\/},  {\it \bibinfo{volume}{533}\/}, \bibinfo{pages}{178--190}.
\bibitem[{Thomee et~al.(2016)Thomee, Shamma, Friedland, Elizalde, Ni, Poland, Borth \& Li}]{thomee2016yfcc100m}
\bibinfo{author}{Thomee, B.}, \bibinfo{author}{Shamma, D.~A.}, \bibinfo{author}{Friedland, G.}, \bibinfo{author}{Elizalde, B.}, \bibinfo{author}{Ni, K.}, \bibinfo{author}{Poland, D.}, \bibinfo{author}{Borth, D.}, \& \bibinfo{author}{Li, L.-J.} (\bibinfo{year}{2016}).
\newblock \bibinfo{title}{Yfcc100m: The new data in multimedia research}.
\newblock {\it \bibinfo{journal}{Communications of the ACM}\/},  {\it \bibinfo{volume}{59}\/}, \bibinfo{pages}{64--73}.
\bibitem[{Torres \& Valdiviezo-Diaz(2020)}]{torres2020hashtags}
\bibinfo{author}{Torres, L.~P.}, \& \bibinfo{author}{Valdiviezo-Diaz, P.} (\bibinfo{year}{2020}).
\newblock \bibinfo{title}{Hashtags recommendations in twitter based on collaborative filtering}.
\newblock In {\it \bibinfo{booktitle}{2020 International Conference of Digital Transformation and Innovation Technology (Incodtrin)}\/} (pp. \bibinfo{pages}{38--44}).
\newblock \bibinfo{organization}{IEEE}.
\bibitem[{Vaswani et~al.(2017)Vaswani, Shazeer, Parmar, Uszkoreit, Jones, Gomez, Kaiser \& Polosukhin}]{vaswani2017attention}
\bibinfo{author}{Vaswani, A.}, \bibinfo{author}{Shazeer, N.}, \bibinfo{author}{Parmar, N.}, \bibinfo{author}{Uszkoreit, J.}, \bibinfo{author}{Jones, L.}, \bibinfo{author}{Gomez, A.~N.}, \bibinfo{author}{Kaiser, {\L}.}, \& \bibinfo{author}{Polosukhin, I.} (\bibinfo{year}{2017}).
\newblock \bibinfo{title}{Attention is all you need}.
\newblock {\it \bibinfo{journal}{Advances in neural information processing systems}\/},  {\it \bibinfo{volume}{30}\/}.
\bibitem[{Wang et~al.(2021)Wang, Wei, Yin, Wu, Song \& Nie}]{wang2021dualgnn}
\bibinfo{author}{Wang, Q.}, \bibinfo{author}{Wei, Y.}, \bibinfo{author}{Yin, J.}, \bibinfo{author}{Wu, J.}, \bibinfo{author}{Song, X.}, \& \bibinfo{author}{Nie, L.} (\bibinfo{year}{2021}).
\newblock \bibinfo{title}{Dualgnn: Dual graph neural network for multimedia recommendation}.
\newblock {\it \bibinfo{journal}{IEEE Transactions on Multimedia}\/}, .
\bibitem[{Wei et~al.(2019)Wei, Cheng, Yu, Zhao, Zhu \& Nie}]{wei2019personalized}
\bibinfo{author}{Wei, Y.}, \bibinfo{author}{Cheng, Z.}, \bibinfo{author}{Yu, X.}, \bibinfo{author}{Zhao, Z.}, \bibinfo{author}{Zhu, L.}, \& \bibinfo{author}{Nie, L.} (\bibinfo{year}{2019}).
\newblock \bibinfo{title}{Personalized hashtag recommendation for micro-videos}.
\newblock In {\it \bibinfo{booktitle}{Proceedings of the 27th ACM International Conference on Multimedia}\/} (pp. \bibinfo{pages}{1446--1454}).
\bibitem[{Yang et~al.(2020{\natexlab{a}})Yang, Wang \& Jiang}]{yang2020sentiment}
\bibinfo{author}{Yang, C.}, \bibinfo{author}{Wang, X.}, \& \bibinfo{author}{Jiang, B.} (\bibinfo{year}{2020}{\natexlab{a}}).
\newblock \bibinfo{title}{Sentiment enhanced multi-modal hashtag recommendation for micro-videos}.
\newblock {\it \bibinfo{journal}{IEEE Access}\/},  {\it \bibinfo{volume}{8}\/}, \bibinfo{pages}{78252--78264}.
\bibitem[{Yang et~al.(2020{\natexlab{b}})Yang, Wu, Li, Li, Gu, Deng \& Wu}]{yang2020amnn}
\bibinfo{author}{Yang, Q.}, \bibinfo{author}{Wu, G.}, \bibinfo{author}{Li, Y.}, \bibinfo{author}{Li, R.}, \bibinfo{author}{Gu, X.}, \bibinfo{author}{Deng, H.}, \& \bibinfo{author}{Wu, J.} (\bibinfo{year}{2020}{\natexlab{b}}).
\newblock \bibinfo{title}{Amnn: Attention-based multimodal neural network model for hashtag recommendation}.
\newblock {\it \bibinfo{journal}{IEEE Transactions on Computational Social Systems}\/},  {\it \bibinfo{volume}{7}\/}, \bibinfo{pages}{768--779}.
\bibitem[{Yong-jun et~al.(2022)Yong-jun, Huan-jing \& Li-jun}]{yong2022implicit}
\bibinfo{author}{Yong-jun, W.}, \bibinfo{author}{Huan-jing, H.}, \& \bibinfo{author}{Li-jun, T.} (\bibinfo{year}{2022}).
\newblock \bibinfo{title}{Implicit tag collaborative filtering recommendation algorithm based on lda}.
\newblock {\it \bibinfo{journal}{Computer and Modernization}\/},  (p.~\bibinfo{pages}{53}).
\bibitem[{Yu et~al.(2023)Yu, Yu, Liang, Mao, Nie, Huang, Khabsa, Fung \& Wang}]{yu2023generating}
\bibinfo{author}{Yu, T.}, \bibinfo{author}{Yu, H.}, \bibinfo{author}{Liang, D.}, \bibinfo{author}{Mao, Y.}, \bibinfo{author}{Nie, S.}, \bibinfo{author}{Huang, P.-Y.}, \bibinfo{author}{Khabsa, M.}, \bibinfo{author}{Fung, P.}, \& \bibinfo{author}{Wang, Y.-C.} (\bibinfo{year}{2023}).
\newblock \bibinfo{title}{Generating hashtags for short-form videos with guided signals}.
\newblock In {\it \bibinfo{booktitle}{Proceedings of the 61st Annual Meeting of the Association for Computational Linguistics (Volume 1: Long Papers)}\/} (pp. \bibinfo{pages}{9482--9495}).
\bibitem[{Zhang et~al.(2019)Zhang, Yao, Xu, Tong, Yan \& Lu}]{zhang2019hashtag}
\bibinfo{author}{Zhang, S.}, \bibinfo{author}{Yao, Y.}, \bibinfo{author}{Xu, F.}, \bibinfo{author}{Tong, H.}, \bibinfo{author}{Yan, X.}, \& \bibinfo{author}{Lu, J.} (\bibinfo{year}{2019}).
\newblock \bibinfo{title}{Hashtag recommendation for photo sharing services}.
\newblock In {\it \bibinfo{booktitle}{Proceedings of the AAAI Conference on Artificial Intelligence}\/} (pp. \bibinfo{pages}{5805--5812}).
\newblock volume~\bibinfo{volume}{33}.
\bibitem[{Zhang et~al.(2023)Zhang, Malkov, Florez, Park, McWilliams, Han \& El-Kishky}]{zhang2023twhin}
\bibinfo{author}{Zhang, X.}, \bibinfo{author}{Malkov, Y.}, \bibinfo{author}{Florez, O.}, \bibinfo{author}{Park, S.}, \bibinfo{author}{McWilliams, B.}, \bibinfo{author}{Han, J.}, \& \bibinfo{author}{El-Kishky, A.} (\bibinfo{year}{2023}).
\newblock \bibinfo{title}{Twhin-bert: A socially-enriched pre-trained language model for multilingual tweet representations at twitter}.
\newblock In {\it \bibinfo{booktitle}{Proceedings of the 29th ACM SIGKDD Conference on Knowledge Discovery and Data Mining}\/} (pp. \bibinfo{pages}{5597--5607}).
\bibitem[{Zheng et~al.(2022)Zheng, Chen, Du \& Song}]{zheng2022multiview}
\bibinfo{author}{Zheng, J.}, \bibinfo{author}{Chen, S.}, \bibinfo{author}{Du, Y.}, \& \bibinfo{author}{Song, P.} (\bibinfo{year}{2022}).
\newblock \bibinfo{title}{A multiview graph collaborative filtering by incorporating homogeneous and heterogeneous signals}.
\newblock {\it \bibinfo{journal}{Information Processing \& Management}\/},  {\it \bibinfo{volume}{59}\/}, \bibinfo{pages}{103072}.
\bibitem[{Zhou et~al.(2022)Zhou, Zhuang, Ren, Chen, Yu, Lou \& Wang}]{zhou2022hybrid}
\bibinfo{author}{Zhou, Q.}, \bibinfo{author}{Zhuang, W.}, \bibinfo{author}{Ren, H.}, \bibinfo{author}{Chen, Y.}, \bibinfo{author}{Yu, B.}, \bibinfo{author}{Lou, J.}, \& \bibinfo{author}{Wang, Y.} (\bibinfo{year}{2022}).
\newblock \bibinfo{title}{Hybrid collaborative filtering model for consumer dynamic service recommendation based on mobile cloud information system}.
\newblock {\it \bibinfo{journal}{Information Processing \& Management}\/},  {\it \bibinfo{volume}{59}\/}, \bibinfo{pages}{102871}.
\bibitem[{Zhu et~al.(2019)Zhu, Yang \& Li}]{zhu2019learning}
\bibinfo{author}{Zhu, R.}, \bibinfo{author}{Yang, D.}, \& \bibinfo{author}{Li, Y.} (\bibinfo{year}{2019}).
\newblock \bibinfo{title}{Learning improved semantic representations with tree-structured lstm for hashtag recommendation: An experimental study}.
\newblock {\it \bibinfo{journal}{Information}\/},  {\it \bibinfo{volume}{10}\/}, \bibinfo{pages}{127}.

\end{thebibliography}

\end{document}